\documentclass[prd,aps,showpacs,superscriptaddress,nofootinbib,amsmath,amssymb]{revtex4}

 \usepackage{graphicx}
\usepackage{dcolumn}
\usepackage{bm}
\usepackage{lipsum}
\begin{document}


\title{Nuclear medium effects in structure functions of nucleon at moderate $Q^2$}
\author{H. Haider}
\author{F. Zaidi\footnote{Corresponding author: zaidi.physics@gmail.com}}
\author{M. Sajjad Athar}
\author{S. K. Singh}
\affiliation{Department of Physics, Aligarh Muslim University, Aligarh - 202 002, India}
\author{I. \surname{Ruiz Simo}}
\affiliation{Departamento de F\'{\i}sica At\'omica, Molecular y Nuclear,
and Instituto de F\'{\i}sica Te\'orica y Computacional Carlos I,
Universidad de Granada, Granada 18071, Spain}
\begin{abstract}
Recent experiments performed on inclusive electron scattering from nuclear targets have measured the nucleon 
electromagnetic structure functions $F_1(x,Q^2)$, $F_2(x,Q^2)$ and 
$F_L(x,Q^2)$ in  $^{12}C$, $^{27}Al$, $^{56}Fe$ and $^{64}Cu$ nuclei. The measurements have been done in the
energy region of $1~GeV^2 < W^2 < 4 ~GeV^2$ and 
$Q^2$ region of $0.5~GeV^2 < Q^2 < 4.5~GeV^2$. We have calculated nuclear medium effects in these structure 
functions arising due to the Fermi motion,
binding energy, nucleon correlations, mesonic contributions from pion and rho mesons and shadowing effects.
The calculations are performed in 
a local density approximation using a
relativistic nucleon spectral function which includes nucleon correlations.
The numerical results are compared with the recent experimental data from JLab and also with 
some earlier experiments.
\end{abstract}
\maketitle
\section{Introduction}
Charged lepton induced processes in the deep inelastic scattering region are used to probe quark and gluon structures of 
nucleons and nuclei. The inclusive lepton scattering from nucleon and nuclear targets is an important tool to study 
the nucleon structure and its modification in nuclear medium. 
The early results from the experiments performed at SLAC in the kinematic region of 
high energy transfer($\nu$) and high momentum transfer($Q^2$) corresponding to Deep 
Inelastic Scattering (DIS) exhibited remarkable phenomenon of Bjorken scaling~\cite{Kendall:1991np}. 
In Bjorken scaling, the nucleon structure functions in the asymptotic region 
of very high $Q^2$ are found to be independent of $Q^2$ and depend only upon single dimensionless
variable $x(=\frac{Q^2}{2M\nu})$ instead of otherwise independent variables $Q^2$ and $\nu$. 
This $Q^2$ independence of nucleon structure functions led to the first evidence that 
nucleons consist of structureless constituents identified as quarks and gluons. Furthermore, the 
 lepton-nucleon cross sections were found to be incoherent sum of elastic lepton scattering
 cross section from these structureless constituents confirming the asymptotic freedom predicted by QCD. 
 At lower values of $Q^2$ and x, the nucleon structure functions exhibit $Q^2$ dependence 
 which is attributed to the violation of Bjorken scaling due to higher twist correction in QCD~\cite{Jaffe:1981td,Donnachie:1980iy,Nachtmann:1981qy}. 
 These higher twist corrections arise due to quark-quark and quark-gluon 
interactions in the nucleon which play an important role when we move to lower $Q^2$ from the asymptotic region in $Q^2$. 
 The nucleon structure functions in the scaling region of high $Q^2$, and their $Q^2$ evolution to lower $Q^2$ region are described rather well 
 using the methods of 
perturbative QCD and evolution equation of DGLAP~\cite{Altarelli:1977zs}.

In the case of deep inelastic lepton scattering from nuclear targets,
the nuclear structure functions per nucleon are found to be quite different from the 
nucleon structure functions as discovered by the EMC effect first observed at CERN ~\cite{emc83}
 and confirmed by many experiments thereafter~\cite{e139_old}-\cite{Arrington:2012ax}. 
 These modifications in the nucleon structure function are due to the 
nuclear medium effects like Fermi motion,
binding energy, nucleon correlations, mesonic contributions, etc. These are in addition to higher twist effects in QCD 
 for nucleons and mesons in nuclear medium~\cite{Miller:2000ta,Castorina:2004ye,Zhang:2014dya}.

The nuclear medium effects in the DIS region are divided
into four parts which are broadly identified in terms of Bjorken scaling variable $x$. These are
(i) the shadowing effect which is effective in the low values of $x$($<0.1$). It has been found that in this 
region {\it R}(=$\frac{\sigma^A}{\sigma^D}$) gets
suppressed, and the suppression increases with the increase 
in the nucleon number $A$, 
(ii) the anti-shadowing effect which is effective in the region $0.1 < x <0.3$, where there is slight enhancement in the ratio {\it R} which
has been found to be independent 
of the nucleon number $A$. Shadowing and anti-shadowing effects are attributed respectively due to the constructive and destructive 
interference of amplitudes arising from the multiple scattering 
of quarks inside the nucleus, (iii) the EMC effect which is a large suppression in a wide range of $x$($0.3 < x <0.8$), and is broadly
understood as due to the modification of 
nucleon structure functions in nuclei, and 
(iv) the binding energy and the Fermi motion effect which is effective for $x>0.8$ and this arises due to the fact that the nucleons in a nucleus are 
moving with an average momentum $p \le p_F$, where $p_F$ is the Fermi momentum.

In the DIS region, phenomenologically~\cite{Global_Fits_1}-~\cite{Kovarik:2012zz} as well as 
theoretically~\cite{Akulinichev1985}-\cite{Armesto:2006ph} various attempts have been made to 
understand the nuclear medium effects. Phenomenologically the studies have been made to obtain a nuclear correction factor 
by doing the analysis of the experimental data on charged lepton-nucleus scattering, (anti)neutrino nucleus scattering,
pion-nucleus scattering, 
 proton-nucleus scattering,
 Drell-Yan processes, etc.
 Theoretically many models have been proposed to study these effects on the basis of 
 nuclear binding, nuclear medium modification including short range correlations in nuclei~\cite{Akulinichev1985}-\cite{Malace:2014uea}, 
 pion excess in nuclei~\cite{Bickerstaff:1989ch, Kulagin1989, marco1996, Ericson:1983um, Bickerstaff:1985mp, Berger1987}, 
 multi-quark clusters~\cite{Jaffe:1982rr, Mineo:2003vc, Cloet:2005rt}, 
 dynamical rescaling~\cite{Nachtmann:1983py, Close:1983tn}, nuclear shadowing~\cite{emcreview_1, Armesto:2006ph}, 
 etc. In spite of these efforts a comprehensive understanding 
 of the nuclear modifications of nucleon structure functions valid for the 
  entire region of $x$ is still lacking~\cite{emcreview_2,Geesaman:1995,Hen:2013oha, Piller:1999wx}.

  Recently JLab~\cite{Mamyan:2012th} has performed experiments using continuous electron beam facility with 
  energies in the range of approximately 2-6 $GeV$, and precise measurements have 
  been performed for the nucleon structure functions $F_{1}^N(x,Q^2)$, $F_{2}^N(x,Q^2)$ and longitudinal structure function $F_{L}^{N}(x,Q^2)$
 in the energy region of $1~GeV^2 < W^2 < 4~ GeV^2$ corresponding to low and moderate $Q^2$ in the region of
 $0.5~GeV^2 < Q^2 < 4.5~GeV^2$ on several nuclear targets like $^{12}C$, $^{27}Al$, $^{56}Fe$, $^{64}Cu$, $^{119}Sn$, etc.
  The modification of structure function in nuclear medium has also been studied earlier by SLAC~\cite{e139},  NMC~\cite{nmcca2d}, BCDMS~\cite{Benvenuti:1987zj}, etc.
  collaborations, in some of these nuclear targets as well as in a few
  other nuclear targets like $^{108}Ag$, $^{197}Au$, $^{208}Pb$, etc. JLab also plans to upgrade electron beam energy
  to 12 $GeV$~\cite{jlab,jlabupdate} and measure nucleon structure functions at low and moderate $Q^2$ in the region of $W$
  relevant for the study of quark-hadron duality in nuclei. It is, therefore, important that nuclear medium effects are theoretically
  well understood in the deep inelastic region as well as in the resonance production region.
  
  Motivated by the recent experimental results~\cite{Mamyan:2012th}, we have performed a theoretical study to understand
  nuclear medium effect in nuclear structure functions $F_{1~N}^A(x,Q^2)$, $F_{2~N}^A(x,Q^2)$ as well as longitudinal 
  structure function $F_{L~N}^A(x,Q^2)$ in nuclei for 
  several nuclear targets like $^{12}C$, $^{27}Al$, $^{56}Fe$, $^{64}Cu$, $^{119}Sn$, $^{197}Au$ and $^{208}Pb$ in the region of moderate $Q^2$.
  Our numerical results are compared with the available JLab data~\cite{Mamyan:2012th} and also with the available results from earlier experiments 
  performed by NMC~\cite{nmcca2d} collaboration. This study is a continuation of our previous study of 
  nuclear medium effects in the extraction of 
  electromagnetic and weak structure functions  $F_{2;~A}^{EM}(x,Q^2)$~\cite{sajjadnpa}, $F_{2,3;~A}^{Weak}(x,Q^2)$\cite{sajjadplb,prc84,prc85}, 
  Paschos-Wolfenstein relation~\cite{prc87} and Parity Violating asymmetry in deep inelastic polarized 
  electron scattering~\cite{Haider:2014iia} from nuclear targets. 
  In the present paper, we have obtained the expressions of $F_{1~N}^A(x,Q^2)$ and $F_{2~N}^A(x,Q^2)$ separately without using the Callan-Gross relation. 
  Therefore, these results can also provide a theoretical framework for testing the validity of Callan-Gross 
  relation in nuclei. 
  
  The present study has been performed with a microscopic model which uses
  relativistic nucleon spectral function to describe target nucleon momentum distribution incorporating 
  Fermi motion, 
binding energy effects and nucleon correlations in a field theoretical model. The spectral function that describes the energy and
momentum distribution 
of the nucleons in nuclei is obtained by using the Lehmann's representation for the relativistic nucleon propagator 
and nuclear many body theory is used to calculate it for an interacting Fermi sea in nuclear matter~\cite{FernandezdeCordoba:1991wf}.
A local density approximation is then applied to translate these results 
to a finite nucleus. Furthermore, we have considered the contributions of the pion and rho meson clouds in a many body field theoretical approach 
based on Refs.~\cite{marco1996,GarciaRecio:1994cn}. Due to the fact that JLab data~\cite{Mamyan:2012th},~\cite{jlab} have been taken in 
a region of relatively low $ Q^2 $  
 we have not assumed the Bjorken limit. We have incorporated Target Mass Correction (TMC) following Ref.~\cite{Schienbein:2007gr}. 
 This is effective at 
 moderate $Q^2$ and high $x$. The 
 calculations are 
 performed at the leading order(LO) as well as next to the leading order (NLO). The nucleon Parton Distribution Functions(PDFs) 
 have been taken from 
 the works of CTEQ group~\cite{cteq}. The NLO evolution of the deep inelastic structure functions has been taken from the works of 
 Vermaseren et al.~\cite{Vermaseren} and van Neerven and Vogt~\cite{Neerven}. 
 In the case of pions we have taken the pionic parton distribution functions given by Gluck et al.~\cite{Gluck:1991ey,Gluck}. 
 For the rho mesons, we have applied the same PDFs as for the pions as in Ref.~\cite{marco1996}.  We have also considered the effect of 
 shadowing following Ref.~\cite{Petti}. 

The plan of presentation is the following. In Sect.~\ref{DIS_lepton_nucleon} some basic formalism for the inclusive lepton-nucleon scattering has been introduced. 
In Sect.~\ref{sec:NE} the various nuclear medium effects have been discussed in brief, and in 
Sect.~\ref{sec:RE} the numerical results are presented. In Sect.~\ref{sec:Summary}, we 
conclude our findings.
\begin{figure}
\begin{center}
\includegraphics{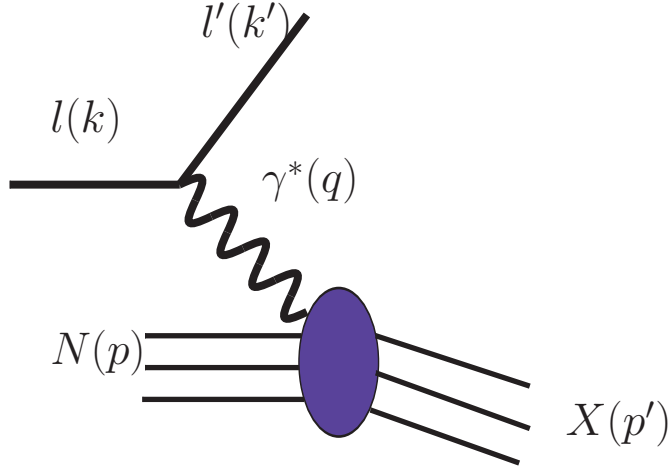}
\caption{Feynman diagram for the virtual photon induced deep inelastic lepton-nucleon scattering. $l(l^\prime)$ is a lepton of four momenta $k(k^\prime)$,
$N$ is a nucleon with four momentum $p$ and $X$ is the jet of hadrons with four momentum $p^\prime$.}
\label{fg:fig1}
\end{center}
\end{figure}
\section{Deep inelastic lepton-nucleon scattering } \label{DIS_lepton_nucleon}
The double differential cross section for the reaction (depicted in Fig.\ref{fg:fig1})
\begin{equation} 	\label{reaction}
l(k) + N(p) \rightarrow l(k^\prime) + X(p^\prime),~l=~e^-,~\mu^-,
\end{equation}
 in the Lab frame is written as
\begin{equation}\label{eN}
\frac{d^2 \sigma^N}{d\Omega_l dE_l^{\prime}} = \frac{\alpha^2}{q^4} \; 
\; \frac{|\bf k'|}{|\bf k|} \;L_{\mu \nu} \; W^{\mu \nu}_N,
\end{equation}
where $q=k-k^\prime$ is the four momentum transfer, $q^2(=-4E_lE_l^{\prime}sin^2\frac{\theta}{2})~\le~0$, and $\alpha = e^2 / 4 \pi$. $L_{\mu \nu}$ is the spin averaged leptonic tensor given by
\begin{equation}\label{leptonictensor}
L_{\mu \nu} = 2 (k_{\mu} k'_{\nu} +  k'_{\mu} k_{\nu} - k\cdot k^\prime g_{\mu \nu})
\end{equation}
where $\Omega_l, E'_l$ refer to the outgoing lepton. The hadronic tensor $W^{\mu \nu}_N$ is defined 
in terms of $W_{1}^{N}, ~W_{2}^{N}$ structure functions of the nucleon,
\begin{equation}\label{nucleonht}
W^{\mu \nu}_N = 
\left( \frac{q^{\mu} q^{\nu}}{q^2} - g^{\mu \nu} \right) \;
W_{1}^N + \left( p^{\mu} - \frac{p . q}{q^2} \; q^{\mu} \right)
\left( p^{\nu} - \frac{p . q}{q^2} \; q^{\nu} \right)
\frac{W_{2}^N}{M^2}
\end{equation}
with $M$ as the mass of nucleon.

In terms of the Bjorken variable $x=\frac{Q^2}{2M\nu}=\frac{Q^2}{2M (E_l - E_l^{\prime})}$ and $y=\frac{\nu}{E_l}$, where $Q^2=-q^2$ and $\nu$ is the energy transfer($=E_l-E_l^\prime$),  the differential cross section is given by
\begin{eqnarray}\label{diff_dxdy}
\frac{d^2 \sigma^N}{d x d y}&=&
\frac{8 M E_l \pi \alpha^2 }{Q^4}
\left\{xy^2 F_{1}^N(x, Q^2)
+ \left(1-y-\frac{xyM}{2 E_l}\right) F_{2}^N(x, Q^2)
\right\}\,.
\end{eqnarray}
where $F_1^N(x,Q^2)=MW_1^N(\nu,Q^2)$ and $F_2^N(x,Q^2)=\nu W_2^N(\nu,Q^2)$, are the dimensionless structure functions. In the Bjorken 
limit, i.e. $Q^2 \rightarrow \infty$, $\nu \rightarrow \infty$, $x$ finite, the structure functions $F_{i=1,2}^N(x, Q^2)$  depend only 
on the  variable $x$  and satisfy the
Callan-Gross relation~\cite{Callan} given by $2xF_1^N(x,Q^2)=F_2^N(x,Q^2)$. 

The total cross section for polarized photon (helicity $\lambda$) interacting with unpolarized proton is expressed as~\cite{Renton}
\begin{equation}
 \sigma_\lambda=\frac{4\pi^2\alpha}{K} {\epsilon_\lambda^\mu}^\ast {\epsilon_\lambda^\nu} W_{\mu\nu}
\end{equation}
where $\epsilon_\lambda^\mu$ is the polarization vector of the photon with $\lambda=\pm 1, 0$; $K~=~\frac{W^2 - M^2}{2M}$, 
$W$ is the invariant energy of the virtual photon-proton system i.e $\sqrt{W^2}=\sqrt{M^2+2M(E_l~-~E_l^\prime)-Q^2}$ and 
 $W_{\mu\nu}$ is given by Eq.(\ref{nucleonht}). Using the above expression, one may obtain the transverse and longitudinal cross sections as
 \begin{eqnarray}\label{eq:LT}
 \sigma_T&=&\frac{4\pi^2\alpha}{K}~W_1^{N}(\nu,~q^2)\nonumber\\
 \sigma_L&=&\frac{4\pi^2\alpha}{K}~\left[\left(1~+~\frac{\nu^2}{Q^2}\right)~W_2^{N}(\nu,~q^2)~-~ W_1^{N}(\nu,~q^2) \right]~~
 \end{eqnarray}
 
 Now using the above expressions (i.e. Eq.(\ref{eq:LT})), $W_1^{N}(\nu,~q^2)$ and $W_2^{N}(\nu,~q^2)$ may be written in terms of $\sigma_L$ and $\sigma_T$, 
 and the differential scattering cross section is written as
\begin{eqnarray}
\frac{d^2 \sigma^N}{d\Omega_l dE_l^{\prime}} &= &\Gamma_T\left[\sigma_T(x,Q^2) +
\epsilon \sigma_L(x,Q^2)\right]  = \Gamma_T \sigma_T(x,Q^2) \left[1 +\epsilon {R^\prime(x,Q^2)}\right]\,,
\label{eq:cs1}
\end{eqnarray}
where $\Gamma_T$ is the transverse virtual photon flux 
$\left( = \alpha E^{\prime}_l/[4 \pi^2 Q^2 M E_l (1 - \epsilon)]\right)$, $R^\prime=\sigma_L/\sigma_T$, 
$\epsilon$ is the virtual photon polarization parameter given by
\begin{eqnarray}
\epsilon & = & \frac{1}{\left[1 + 2(1+\frac{\nu^2}{Q^2}) {\tan}^2 \frac{\theta_{Lab}}{2}\right]} \simeq \frac{1-y-\frac{M^2 x^2 y^2}{Q^2}}
{1-y+\frac{y^2}{2}+\frac{M^2 x^2 y^2}{Q^2}}\,.
\label{eq:epsilon}
\end{eqnarray}
 $\theta_{Lab}$ is the Lab scattering angle.
 
Using the above expressions for $\Gamma_T$ and $\epsilon$, one may re-write Eq.(\ref{eN}) as
\begin{eqnarray}
\frac{d^2 \sigma^N}{d\Omega_l dE_l^{\prime}} & = & \Gamma_T \frac{4{\pi}^2 
\alpha}{x(W^2-M^2)}  \left[ 2xF_1^N(x,Q^2)  + \epsilon \left( (1+\frac{4M^2 x^2}{Q^2}) F_2^N(x,Q^2)-2xF_1^N(x,Q^2) \right) \right].~~~
\label{eq:cs2}
\end{eqnarray}

The expression for the longitudinal structure 
function $F_L^N(x,Q^2)$ is defined as
\begin{equation}
F_L^N(x,Q^2)= (1+\frac{4M^2 x^2}{Q^2}) F_2^N(x,Q^2)-2xF_1^N(x,Q^2)
\label{eq:fl}
\end{equation}
where $F_1^N(x,Q^2)$ is purely transverse and $F_2^N(x,Q^2)$ is a mixture of both. 

The nucleon structure functions are determined in terms of parton distribution functions for quarks and anti-quarks.
In this work, for the nucleons we work at the leading order(LO) and next to leading order(NLO) and used CTEQ6.6~\cite{cteq} nucleon 
Parton Distribution Functions (PDFs). For $Q^2$ evolution we have used the model of Refs.~\cite{Vermaseren,Neerven,Moch:2004xu}. It may be pointed out 
that we have also obtained the results at next to next to leading order i.e. NNLO (the results not shown here) and found the change between NLO and NNLO results to be less than
 half a percent.

\section{Nuclear effects}\label{sec:NE}
Now we evaluate the cross section for the reaction given by Eq.(\ref{reaction}) for the lepton scattering taking place with a nucleon 
moving inside the nucleus. The expression of the  cross section given in Eq.(\ref{eN}) is modified as
\begin{equation}\label{eA}
\frac{d^2 \sigma^A}{d\Omega_l dE_l^{\prime}} =~\frac{\alpha^2}{q^4} \; \frac{|\bf k'|}{|\bf k|} \;L_{\mu \nu} \; W_A^{\mu \nu},
\end{equation}
where $W_A^{\mu \nu}$ is the nuclear hadronic tensor defined in terms of nuclear hadronic structure functions $W_i^A$(i=1,2) as

\begin{equation}\label{nuclearht}
W_A^{\mu \nu} = 
\left( \frac{q^{\mu} q^{\nu}}{q^2} - g^{\mu \nu} \right) \;
W_1^A + \left( p_A^{\mu} - \frac{p_A . q}{q^2} \; q^{\mu} \right)
\left( p_A^{\nu} - \frac{p_A . q}{q^2} \; q^{\nu} \right)
\frac{W_2^A}{M_A^2}
\end{equation}
with $M_A$ is the mass of nucleus.
\begin{figure}
\includegraphics[width=8cm,height=4.8cm]{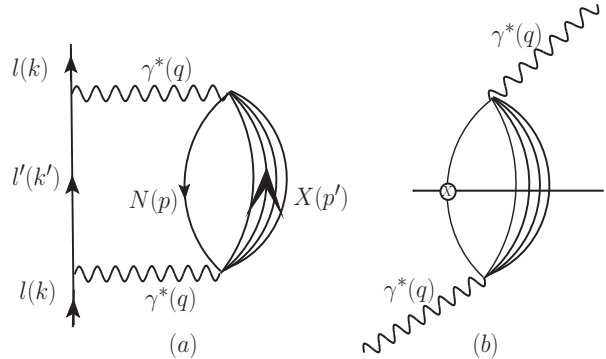}
\caption{(a) Lepton self energy (b) Photon self energy. The imaginary part is calculated by cutting along the horizontal
line and applying the Cutkosky rules while putting the particle on 
mass shell.}
\label{pse}
\end{figure}
In the present formalism the lepton-nucleus DIS cross section is obtained in terms of lepton self energy in the 
 nuclear medium. We evaluate the lepton self energy corresponding to the diagram shown in Fig.\ref{pse}, 
 and the cross section for an element of volume $dV$ in the nucleus is related to the
 probability per unit time ($\Gamma$) of the lepton interacting with the nucleons. $\Gamma dt dS$ provides probability times a differential of 
area(dS) which is nothing but the cross section(d$\sigma$) i.e.
 \begin{eqnarray}\label{defxsec}
d\sigma&=&\Gamma dt dS=\Gamma \frac{dt}{dl}dS dl=\Gamma \frac{1}{v}dV = \Gamma \frac {E_l}{\mid {\bf k} \mid}dV = \Gamma \frac {E_l}{\mid {\bf k} \mid}d^3r,
\end{eqnarray}
where d$l$ is the length of the interaction, $v(=\frac{dl}{dt})$ is the velocity of the incoming lepton and we have used ${\bf k}={\bf v}E_l$.
 
 Also probability per unit time of the lepton interacting with the nucleons in the medium to give the final state is related to the imaginary part of the 
 lepton self energy i.e.
 \begin{equation}\label{pro1}
 -\frac{\Gamma}{2}=\frac{m}{E_l({\bf k})}Im\Sigma
\end{equation}
 Equating  Eq.~\ref{defxsec} and Eq.\ref{pro1}, one gets
\begin{equation}\label{defxsec1}
 d\sigma=\frac{-2m}{E_l({\bf k})} Im \Sigma (k)\frac{E_l({\bf k})}{\mid {\bf k} \mid}d^3r,
\end{equation}
 Thus to get $d\sigma$ for $(l, l')$ scattering on the nucleus given by Eq.~\ref{eA}, we are required to evaluate
imaginary part of lepton self energy $Im \Sigma (k)$. For this, we write the lepton self energy $(\Sigma (k))$ corresponding 
to the diagram shown in Fig.\ref{pse}a by using Feynman rules
\begin{eqnarray}
- i \Sigma (k) = \int \frac{d^4 q}{(2 \pi)^4} \; 
{\bar u}_l ({\bf k}) \; i e \gamma^{\mu} \; i \frac{\not \! k' + m}{k'^{2} -
m^2 + i \epsilon} \; i e \gamma^{\nu} u_l ({\bf k}) \; 
\frac{-i g_{\mu \rho}}{q^2} \; (- i) \; \Pi^{\rho \sigma} (q) \; 
\frac{-i g_{\sigma \nu}}{q^2}
\end{eqnarray}
which for unpolarized leptons can be written as

\begin{equation}
\Sigma (k) = i e^2 \; \int \frac{d^4 q}{(2 \pi)^4} \;
\frac{1}{q^4} \;
\frac{1}{2m} \;
L_{\mu \nu} \; \frac{1}{k'^2 - m^2 + i \epsilon} \; \Pi^{\mu \nu} (q),
\end{equation}
where $\Pi^{\mu \nu} (q)$ the photon self energy and $L_{\mu \nu}$ is the leptonic tensor given by Eq.\ref{leptonictensor}.
Now we shall use the imaginary part of the lepton self energy  i.e. $Im \Sigma (k)$, to obtain the results for the cross section and for this we apply 
Cutkosky rules~\cite{Itzykson}
\begin{equation}\label{cut}
\begin{array}{lll}
\Sigma (k) & \rightarrow & 2 i \; I m \Sigma (k) \\
D (k') & \rightarrow & 2 i \theta (k'^0) \; I m D (k') \; 
\hbox{(boson propagator)} \\
\Pi^{\mu \nu} (q) & \rightarrow & 2 i \theta (q^0) \; I m \Pi^{\mu \nu} (q) \\
G (p) & \rightarrow & 2 i \theta (p^0) \; I m G (p) \; 
\hbox{(fermion propagator)} 
\end{array}
\end{equation}
which leads to
\begin{eqnarray}\label{self-lepton}
2iIm \Sigma (k) &=& ie^2 \int \frac{d^4 q}{(2 \pi)^4} \;
\frac{1}{q^4} \; \frac{1}{2m} \; L_{\mu \nu}\;
2iIm  \frac{1}{k'^{2}-m^2 + i \epsilon}\;2i Im(\Pi^{\mu\nu})\nonumber\\
\Rightarrow Im \Sigma (k) &=& e^2 \int \frac{d^3 q}{(2 \pi)^3} \;
\frac{1}{2E_l}\theta(q^0) \; Im(\Pi^{\mu\nu})  \frac{1}{q^4} \frac{1}{2m} \; L_{\mu \nu}\;
\end{eqnarray}

 Notice from Eq.~\ref{self-lepton}, $\Sigma (k)$ contains photon self energy $\Pi^{\mu\nu}$, which is 
written in terms of nucleon  propagator $G_l$ and meson  propagator $D_j$ and using Feynman rules~\cite{Itzykson} following Fig. \ref{pse} this is given by
\begin{eqnarray}\label{photonse}
\Pi^{\mu \nu} (q)&=& e^2 \int \frac{d^4 p}{(2 \pi)^4} G (p) 
\sum_X \; \sum_{s_p, s_l} {\prod}_{\substack{i = 1}}^{^N} \int \frac{d^4 p'_i}{(2 \pi)^4} \; \prod_{_l} G_l (p'_l)\; \prod_{_j} \; D_j (p'_j)\nonumber \\  
&&  <X | J^{\mu} | H >  <X | J^{\nu} | H >^* (2 \pi)^4  \; \delta^4 (q + p - \sum^N_{i = 1} p'_i),\;\;\;
\end{eqnarray}
where $s_p$ is the spin of the nucleon, $s_i$ is the spin of the fermions in $X$, $<X | J^{\mu} | H >$ is the hadronic current for the initial state nucleon 
to the final state hadrons, index $l$ runs for fermions and index $j$ runs for bosons in the final hadron state $X$.

 To derive the nucleon propagator in nuclear medium G(p) we start with the relativistic free nucleon Dirac propagator $G^{0}(p_{0},{{\bf p}})$ and write 
it in terms of the
contribution from the positive and negative energy components of the nucleon described by the Dirac spinors
$u({\bf p})$ and $v({\bf p})$ using their appropriate normalization ($\bar{u}u=1$), which results~\cite{Itzykson}:
\begin{equation}  \label{prop2}
G^{0}(p_{0},{{\bf p}}) =\frac{M}{E({\bf p})}\left\{\frac{\sum_{r}u_{r}({\bf p})\bar u_{r}({\bf p})}{p_{0}-E({{\bf p}})+i\epsilon}+\frac{\sum_{r}v_{r}(-{\bf p})
\bar v_{r}(-{\bf p})}{p_{0}+E({{\bf p}})-i\epsilon}\right\}
\end{equation}
where M is the mass and $E({\bf p})$ is the relativistic energy($\sqrt{{\bf p}^2+M^2}$) of an on shell nucleon. 
We shall retain only the positive energy contributions as the negative energy 
 contributions are suppressed. 

Moreover, the relativistic nucleon propagator in a non-interacting Fermi sea may be written as
\begin{eqnarray}  \label{prop4}
G^{0}(p_{0},{{\bf p}})&=&\frac{M}{E({{\bf p}})}\left\{\sum_{r}u_{r}({\bf p})\bar u_{r}({\bf p})
\left[\frac{1-n(\bf{p})}{p_{0}-E({{\bf p}})+i\epsilon}+\frac{n(\bf{p})}{p_{0}-E({{\bf p}})-i\epsilon}\right]+\frac{\sum_{r}v_{r}(-{\bf p})\bar v_{r}(-{\bf p})}{p_{0}+E({{\bf p}})-i\epsilon}\right\}
\end{eqnarray}
where $n({{\bf p}})$ is the occupation number of the nucleons in the Fermi sea, $n({{\bf p}})$=1 for p$\le p_{F_N}$ while $n({{\bf p}})$
=0 for p$> p_{F_N}$. 

\begin{figure}
\begin{center}
\includegraphics[width=10cm,height=5cm]{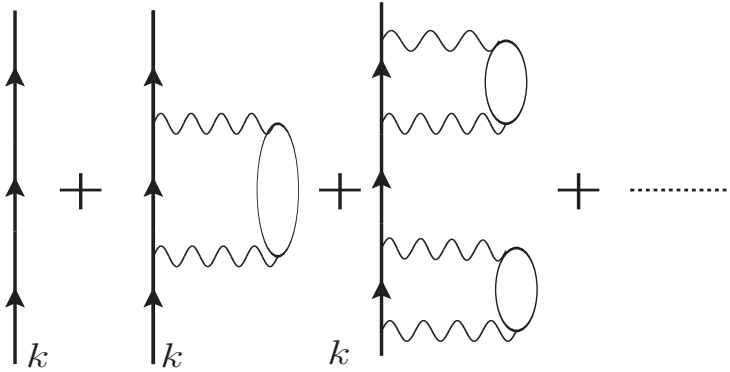}
\caption{Nucleon self-energy in the nuclear medium}
\label{3}
\end{center}
\end{figure}

 In the interacting 
Fermi sea, the relativistic nucleon propagator is written using Dyson series expansion (depicted in Fig.\ref{3}) in terms of 
nucleon self energy $\Sigma^N(p_0,\bf{p})$. This perturbative expansion is summed in a ladder approximation to give~\cite{marco1996}:
\begin{eqnarray}\label{self-sigma}
G(p)&=&\frac{M}{E({\bf p})}\frac{\sum_{r}u_{r}({\bf p})\bar u_{r}({\bf p})}{p_{0}-E({\bf p})}+\frac{M}{E({\bf p})}\frac{\sum_{r}u_{r}({\bf p})\bar
u_{r}({\bf p})}{p_{0}-E({\bf p})}\Sigma^N(p_{0},{\bf p})\frac{M}{E({\bf p})} \frac{\sum_{s}u_{s}({\bf p})\bar u_{s}({\bf p})}{p_{0}-E(p)}+..... \nonumber \\
&=&\frac{M}{E({\bf p})}\sum_{r}\frac{u_{r}({\bf p})\bar u_{r}({\bf p})}{p_{0}-E({\bf p})-\bar u_{r}({\bf p})\Sigma^N(p_{0},{\bf p})u_{r}({\bf p})\frac{M}{{\bf E(p)}}}
\end{eqnarray}

 One may notice from the expression for the nucleon propagator $G(p)$ given in Eq.\ref{self-sigma} that it contains nucleon self energy $\Sigma^N(p_0,{\bf p})$.   
 The nucleon self-energy is written using the techniques of the standard Many-Body Theory~\cite{FernandezdeCordoba:1991wf}. 
 The inputs required for the NN interaction are incorporated by relating them to the experimental elastic NN cross section. 
  Furthermore, RPA-correlation effect is taken into account using the spin-isospin effective 
 interaction as the dominating part of the particle-hole (ph) interaction. Using the modified expression for the nucleon self energy, the imaginary part of 
 it is obtained. Due to RPA effect
 the imaginary part of the nucleon self-energy is quenched specially at low energies and high densities and this also depends on nucleon energy $p_0$ as well as
  nucleon momentum ${\bf p}$ in the interacting Fermi sea. The imaginary part of the nucleon 
 self-energy fulfills the low-density theorem. 
 The real part of the nucleon self energy is obtained by means of dispersion relations using the expressions of the imaginary part.  The Hartree and 
 Fock pieces of the self-energy do not contribute to the imaginary part and therefore cannot be obtained by means of the dispersion relations and are 
 explicitly added. For the real part, the only added piece is the Fock term assuming that the interaction contains spin-isospin excitations having longitudinal 
 and transverse components. This model, of course, misses some pieces of Hartree type, which depend on the density but not on $p_0$ or ${\bf p}$, but its effect is expected 
 to be small in the study of present interest.
 Following Ref.~\cite{marco1996}, the relativistic nucleon propagator G(p) in a nuclear medium is thus expressed as:
\begin{eqnarray}\label{Gp}
G (p) =&& \frac{M}{E({\bf p})} 
\sum_r u_r ({\bf p}) \bar{u}_r({\bf p})
\left[\int^{\mu}_{- \infty} d \, \omega 
\frac{S_h (\omega, {\bf{p}})}{p_0 - \omega - i \eta}
+ \int^{\infty}_{\mu} d \, \omega 
\frac{S_p (\omega, {\bf{p}})}{p_0 - \omega + i \eta}\right]\,,
\end{eqnarray}
where $S_h (\omega, {\bf{p}})$ and $S_p (\omega, {\bf{p}})$ being the hole
and particle spectral functions respectively, which are given by~\cite{marco1996},\cite{FernandezdeCordoba:1991wf}:
\begin{equation}\label{sh}
 S_h(p_0,\mathbf{p})=\frac{1}{\pi}
 \frac{\frac{M}{E(\mathbf{p})}\textrm{Im}\Sigma^N(p_0,\mathbf{p})}{\left(p_0-
 E(\mathbf{p})-\frac{M}{E(\mathbf{p})}\textrm{Re}\Sigma^N(p_0,\mathbf{p})\right)^2+
 \left(\frac{M}{E(\mathbf{p})}\textrm{Im}\Sigma^N(p_0,\mathbf{p})\right)^2}
\end{equation}
for $p_0 \le \mu$
\begin{equation}\label{sp}
 S_p(p_0,\mathbf{p})=-\frac{1}{\pi}
 \frac{\frac{M}{E(\mathbf{p})}\textrm{Im}\Sigma^N(p_0,\mathbf{p})}{\left(p_0-
 E(\mathbf{p})-\frac{M}{E(\mathbf{p})}\textrm{Re}\Sigma^N(p_0,\mathbf{p})\right)^2+
 \left(\frac{M}{E(\mathbf{p})}\textrm{Im}\Sigma^N(p_0,\mathbf{p})\right)^2}
\end{equation}
for $p_0 > \mu$. $\mu$ is the chemical potential given by $\mu=\frac{p_F^2}{2M}~+~Re\Sigma^N\left[\frac{p_F^2}{2M},p_F\right]$and for the present numerical calculations, 
 the expressions for which have been taken from Ref.~\cite{FernandezdeCordoba:1991wf}. 
 A few properties related with spectral function have been 
 presented in Appendix-\ref{app}.  
  In Fig.\ref{spec}, following Ref.~\cite{marco1996}, we have shown $S_h(\omega,\mathbf{p})$ vs $\omega$ (where $\omega=p_0-M$), for $p < p_F$ and $p > p_F$ in $^{12}C$ and $^{56}Fe$ nuclei. 
    It may be observed that for $p < p_F$ the hole spectral function $S_h$ almost mimics a delta function as it corresponds to a Lorentzian distribution with a 
    very narrow width. While for $p > p_F$, $S_h$ is not exactly zero, although very small in magnitude but has a longer range. This behaviour 
    is different from independent particle model where it is exactly zero and this difference arises due to nucleon correlation~\cite{Mahaux:1985zz}.

  Using Eqs.\ref{defxsec1} and \ref{self-lepton}, and performing the momentum space integration one may write the cross section as: 
\begin{equation}\label{dsigma_3}
\frac {d\sigma^A}{d\Omega_l dE_l'}=-\frac{\alpha}{q^4}\frac{|\bf{k^\prime}|}{|\bf {k}|}\frac{1}{(2\pi)^2} L_{\mu\nu} \int  Im \Pi^{\mu\nu}d^{3}r
\end{equation}
At this stage if Eq.(\ref{eA}) and Eq.(\ref{dsigma_3}) are compared it may be inferred that $W_A^{\mu \nu}$ is related with $Im \Pi^{\mu\nu}$ as
\begin{eqnarray}\label{wamunu}
W_A^{\mu \nu}=-\frac{1}{4\pi^2\alpha} \int  Im \Pi^{\mu\nu}d^{3}r
\end{eqnarray}

\begin{figure}
\begin{center} 
 \includegraphics[height=0.3\textheight,width=0.8\textwidth]{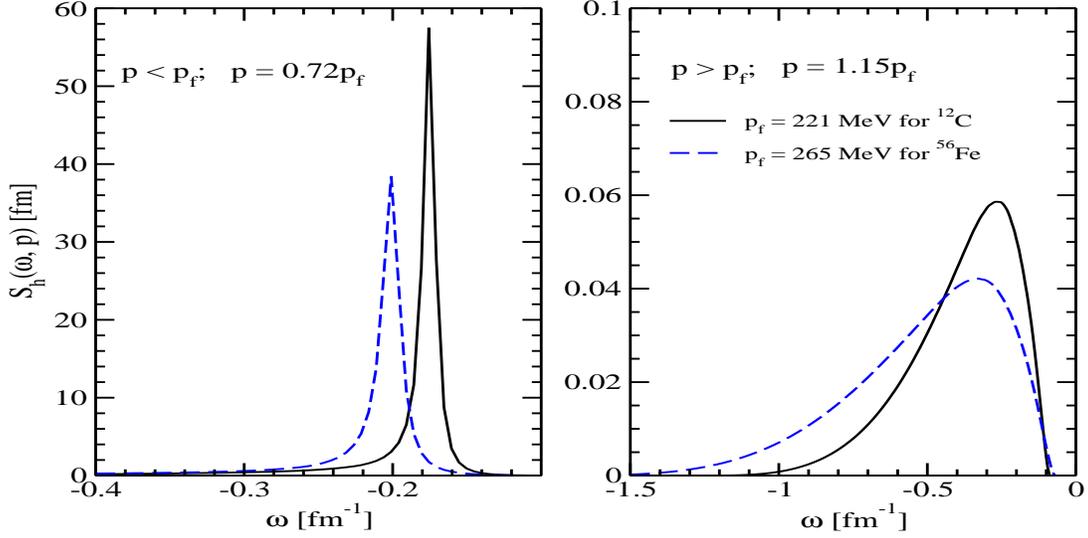}
  \caption{$S_h(\omega,{\bf p})$ vs $\omega$ for $p < p_F$(Left panel) and $p > p_F$(Right panel) in $^{12}C$(solid line) and $^{56}Fe$(dashed line).}
 \label{spec}
 \end{center}
\end{figure}

 Using Eq.(\ref{Gp}) and the expressions for the free nucleon and meson propagators in Eq.(\ref{photonse}), and finally substituting 
 them in Eq.(\ref{wamunu}), we obtain nuclear hadronic tensor for an isospin symmetric nucleus in terms of 
 nucleonic hadronic tensor and spectral function, given by 
\begin{equation}	\label{conv_WA}
W^{\alpha \beta}_{A} = 4 \int \, d^3 r \, \int \frac{d^3 p}{(2 \pi)^3} \, 
\frac{M}{E ({\bf p})} \, \int^{\mu}_{- \infty} d p_0 S_h (p_0, {\bf p}, \rho(r))
W^{\alpha \beta}_{N} (p, q), \,.
\end{equation}
 In this way we have incorporated Fermi motion, Pauli blocking and nucleon correlations through the inclusion of spectral function $S_h(p_0, {\bf p}, \rho(r))$. 
 The normalization of the spectral function and the quantities obtained
 from it are given in Appendix-\ref{app}.
 
 Accordingly the dimensionless nuclear 
structure functions $F_{i=1,2}^A(x,Q^2)$, are defined in terms of $W_{i=1,2}^A(\nu,Q^2)$ as
\begin{eqnarray}\label{relation1}
F_1^A(x,Q^2)&=&M_A~W_{1}^{A}(\nu,Q^2)~\nonumber \\
F_2^A(x,Q^2)&=&\nu_A~W_{2}^{A}(\nu,Q^2)~  {\rm where}\nonumber \\
\nu_A&=&\frac{p_{_A}\cdot q}{M_{_A}}=\frac{p_{0_A} q_{0}}{M_{_A}}=q_{0},~~ p_{_A}^\mu=(M_{_A},\vec 0) {\rm ~and~} M_{_A} {\rm ~is~ the~ mass~ of~ a~ nucleus.}
\end{eqnarray}

Taking the xx component of Eq.(\ref{nucleonht}), we have 
\begin{eqnarray} \label{had-ten1}
W^{N}_{xx} =\left( \frac{q_x q_x}{q^2} - g_{xx} \right) \;W_{1}^N
+ \frac{1}{M^2}\left( p_{x} - \frac{p . q}{q^2} \; q_{x} \right) \left( p_{x} - \frac{p . q}{q^2} \; q_{x} \right)\;W_{2}^N 
\end{eqnarray}
Choosing ${\bf q}$ along the z-axis, we obtain for $q^\mu=(q_0,0,0,q)$ and $p^\mu=(E_N,{\bf p})$,
\begin{eqnarray} \label{had_nucl-ten1}
W^{N}_{xx}(x_N, Q^2)=W_{1}^N(x_N, Q^2) + \frac{1}{M^2}p_{x}^{2} W_{2}^N(x_N, Q^2)
\end{eqnarray}
Similarly taking the xx component of Eq.(\ref{nuclearht}) and using $p_{_A}^\mu=(M_{_A},\vec 0)$, we have 
\begin{eqnarray} \label{had_A-ten1}
W^{A}_{xx}(x_A, Q^2) =W_1^A(x_A, Q^2)=\frac{F_{1}^A(x_A, Q^2)}{AM}
\end{eqnarray}

Using equations Eq.(\ref{had_nucl-ten1}) and Eq.(\ref{had_A-ten1}) in Eq.(\ref{conv_WA}), we have
\begin{eqnarray}	\label{conv_WA1}
F_{_{1~N}}^A(x_A, Q^2) &=& 4AM \int \, d^3 r \, \int \frac{d^3 p}{(2 \pi)^3} \, 
\frac{M}{E ({\bf p})} \, \int^{\mu}_{- \infty} d p_0 S_h(p_0, {\bf p}, \rho(r))~\left[\frac{F_{1}^N(x_N, Q^2)}{M}\right. \nonumber\\
&& \left. + \frac{1}{M^2}{p_x}^2 \frac{F_{2}^N(x_N, Q^2)}{\nu}\right],~~~~
\end{eqnarray}
where $x_N=\frac{Q^2}{2p\cdot q}=\frac{Q^2}{2(p_0q_0-p_zq_z)}$ and $x_A=\frac{x}{A}=\frac{1}{A}\frac{Q^2}{2Mq_0}$.

For nonisoscalar nuclear target the above equation is written as
\begin{eqnarray}	\label{conv_WA1}
F_{_{1~N}}^A(x_A, Q^2) &=& 2\sum_{\tau=p,n} AM \int \, d^3 r \, \int \frac{d^3 p}{(2 \pi)^3} \, 
\frac{M}{E ({\bf p})} \, \int^{\mu}_{- \infty} d p_0 S_h^\tau (p_0, {\bf p}, \rho^\tau(r))~\left[\frac{F_{1}^\tau(x_N, Q^2)}{M}\right. \nonumber\\
&& \left. + \frac{1}{M^2}{p_x}^2 \frac{F_{2}^\tau(x_N, Q^2)}{\nu}\right].~~~
\end{eqnarray}

Similarly taking the  zz component of Eq.(\ref{nucleonht}) and Eq.(\ref{nuclearht}), we obtain
\begin{eqnarray} \label{had_ten11}
W^{N}_{zz}(x_N, Q^2) &=& \frac{q_0^2}{q^2} \;W_{1}^N(x_N, Q^2)~
+ \frac{1}{M^2}\left( \frac{(p_{z} q^2 - {p . q} \; q_{z})^2}{q^4} \right)~W_{2}^N(x_N, Q^2) 
\end{eqnarray}

\begin{eqnarray} \label{had_ten12}
W^{A}_{zz}(x_A, Q^2) &=&\frac{q_0^2}{q^2}\;W_{_{1~N}}^A(x_A, Q^2)~
+ \frac{q_0^2~q_z^2}{q^4}W_{_{2~N}}^A(x_A, Q^2)
\end{eqnarray}
\begin{figure}
\begin{center} 
 \includegraphics[height=0.3\textheight,width=0.8\textwidth]{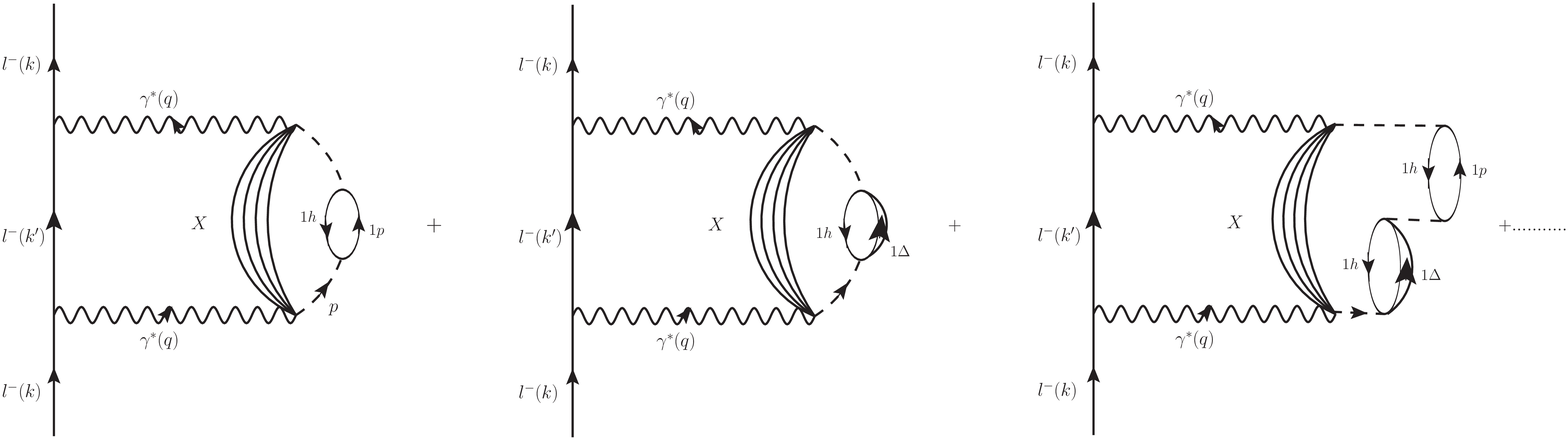}
  \caption{Lepton self energy diagram including particle-hole(1p-1h), delta-hole(1$\Delta$-1h), 1p1h-1$\Delta$1h, etc. excitations.}
 \label{meson}
 \end{center}
\end{figure}

Using equations Eq.(\ref{had_ten11}) and Eq.(\ref{had_ten12}) in Eq.(\ref{conv_WA}), we have
\begin{eqnarray} \label{had_ten3}
\frac{q_0^2}{q^2}\;W_{_{1~N}}^A(x_A, Q^2)~
+ \frac{q_0^2~q_z^2}{q^4}W_{_{2~N}}^A(x_A, Q^2)  &=&  2\sum_{\tau=p,n} \int \, d^3 r \, \int \frac{d^3 p}{(2 \pi)^3} \, 
\frac{M}{E ({\bf p})} \, \int^{\mu}_{- \infty} d p_0 S_h^\tau (p_0, {\bf p}, \rho^\tau(r))\times\nonumber\\ 
&& \left[\frac{q_0^2}{q^2} \;W_{1}^\tau(x_N, Q^2)~
+ \frac{1}{M^2}\left( \frac{(p_{z} q^2 - {p . q} \; q_{z})^2}{q^4} \right)~W_{2}^\tau(x_N, Q^2)\right] 
\end{eqnarray}

 With the expressions of $W_{_{1~N}}^A(x,Q^2)$ and $W_{_{2~N}}^A(x,Q^2)$ in terms of $F_{1~N}^A(x,Q^2)$ and $F_{2~N}^A(x,Q^2)$,
 we obtain from Eq.(\ref{had_ten3}) following expression for $F_{2~N}^A(x,Q^2)$:
 \begin{eqnarray} \label{had_ten151}
F_{_{2~N}}^A(x,Q^2)  &=&  2\sum_{\tau=p,n} \int \, d^3 r \, \int \frac{d^3 p}{(2 \pi)^3} \, 
\frac{M}{E ({\bf p})} \, \int^{\mu}_{- \infty} d p_0 S_h^\tau (p_0, {\bf p}, \rho^\tau(r)) \times\left[\frac{Q^2}{q_z^2}\left( \frac{|{\bf p}|^2~-~p_{z}^2}{2M^2}\right)\right. \nonumber \\
&& \left. +  \frac{(p_0~-~p_z~\gamma)^2}{M^2} \left(\frac{p_z~Q^2}{(p_0~-~p_z~\gamma) q_0 q_z}~+~1\right)^2\right]~\frac{M}{p_0~-~p_z~\gamma} ~F_2^\tau(x,Q^2),       
\end{eqnarray}
where $\gamma=\frac{q_z}{q_0}$.
\begin{figure}
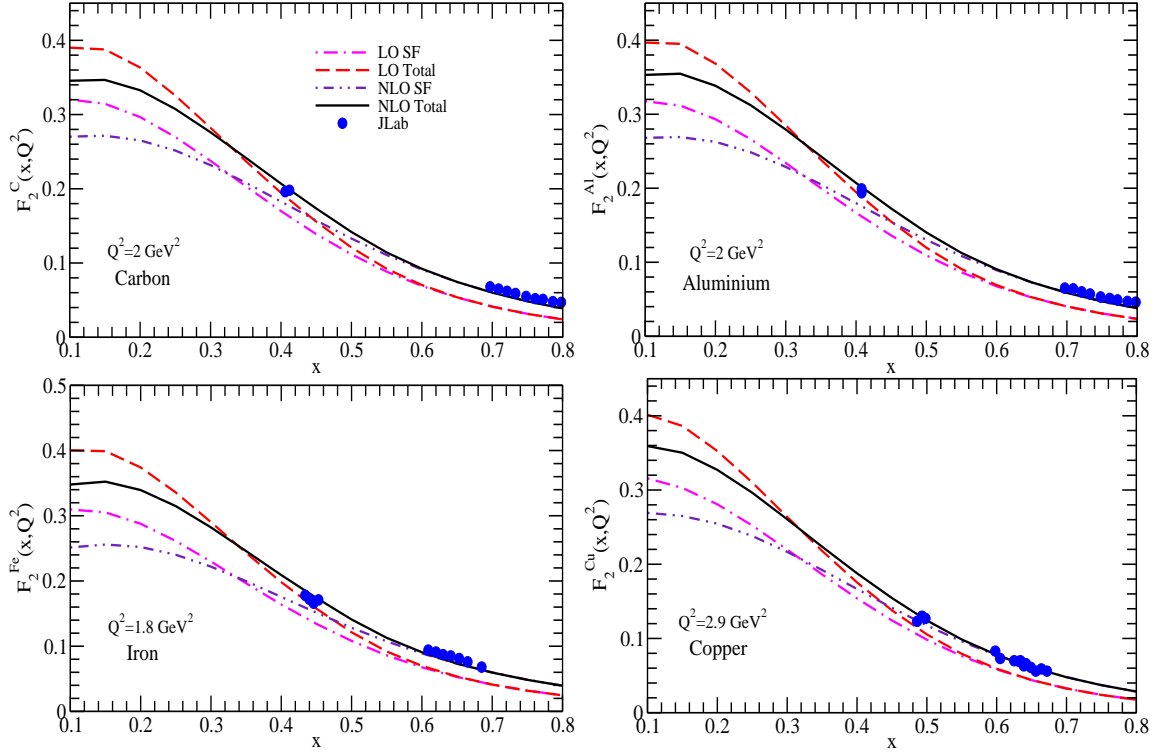

\begin{center} 
 \includegraphics[height=0.21\textheight,width=0.42\textwidth]{f2_lonlo_C.eps}
 \includegraphics[height=0.21\textheight,width=0.42\textwidth]{f2_lonlo_Al.eps}
 \includegraphics[height=0.21\textheight,width=0.42\textwidth]{f2_lonlo_Fe.eps}
 \includegraphics[height=0.21\textheight,width=0.42\textwidth]{f2_lonlo_Cu.eps}
 \caption{$F_{2}^A(x,Q^2)$ vs x at a fixed $Q^2$, in A=$^{12}C,~~^{27}Al,~~^{56}Fe,$ and $^{63}Cu$
 with spectral function(dashed-dotted line) and full calculation(dashed line) at LO and the results with spectral function(dashed-double dotted line)
 and full calculation(solid line) at NLO. Experimental points are the JLab data~\cite{Mamyan:2012th}.}
 \label{FigF2_all}
 \end{center}
\end{figure}

\begin{figure}
\begin{center} 
 \includegraphics[height=0.65\textheight,width=0.85\textwidth]{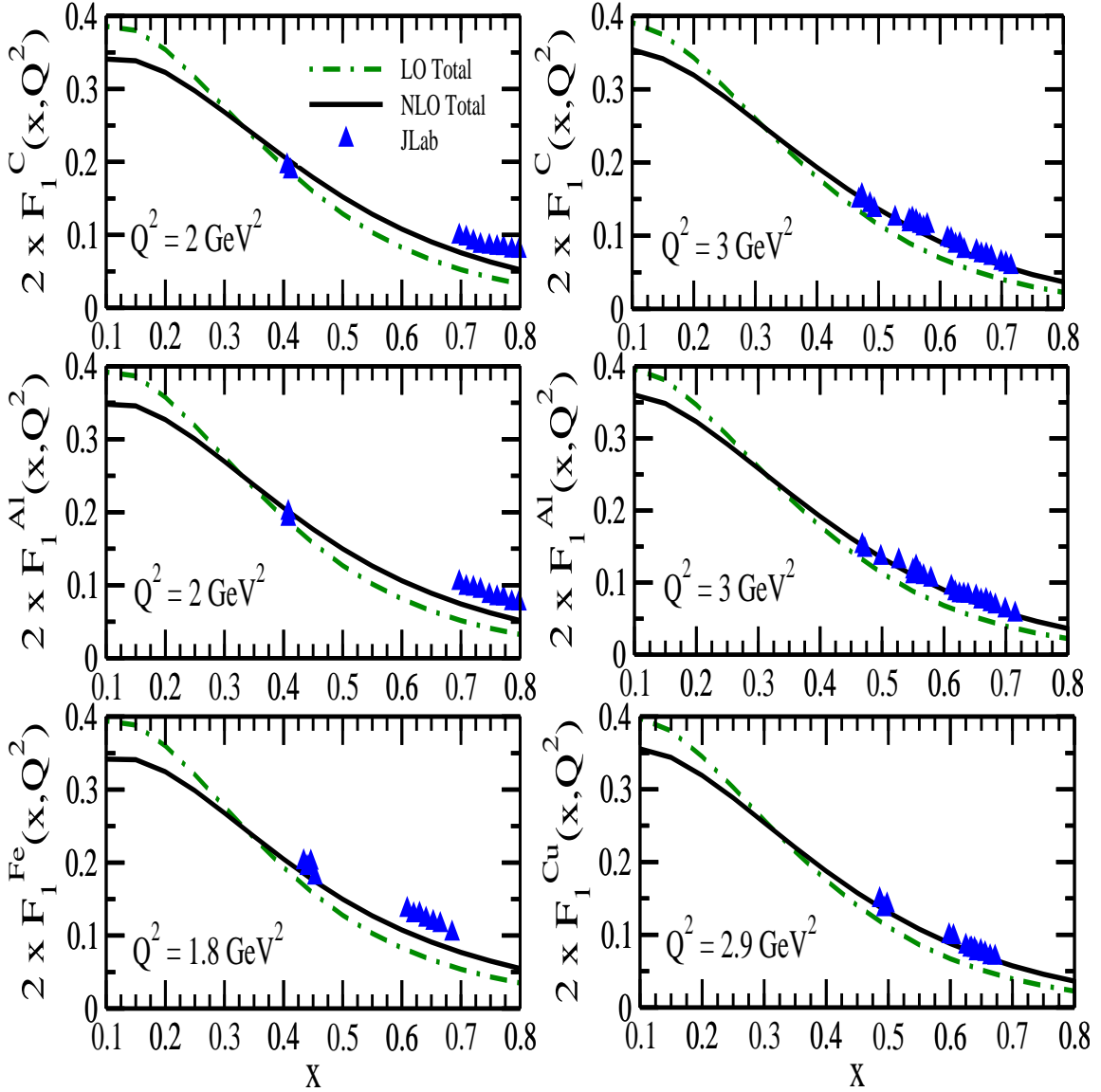}
 \caption{Results of full calculation at LO(LO~Total) and NLO(NLO~Total) for $2xF_1^A(x,Q^2)$(dashed-dotted line: LO~Total and solid line: NLO~Total)
  for A=$^{12}C,~~^{27}Al,~~^{56}Fe,~~^{63}Cu$
  are presented at different $Q^2$. Experimental points are the JLab data~\cite{Mamyan:2012th}.}
  \label{Figq2var_all}
 \end{center}
\end{figure}

Thus for the numerical calculations we shall use Eq.(\ref{conv_WA1}) for $F_{_{1~N}}^A(x,Q^2)$ and Eq.(\ref{had_ten151}) for
$F_{_{2~N}}^A(x,Q^2)$. These numerical results will be labeled as the results 
with spectral functions. Now we include the mesonic contributions coming from pion and rho mesons. We follow the same approach as in the nucleon case. 
However, for completeness, we explicitly present a brief description of this formalism.
\subsection{Mesonic contributions}
 There are virtual mesons associated with each nucleon bound inside the nucleus. These meson clouds get strengthened by the strong 
 attractive nature of nucleon-nucleon interactions. 
 This leads 
 to an increase in the interaction probability of virtual photons with the meson cloud. The effect of meson cloud is more pronounced in heavier 
 nuclear targets and dominate in the intermediate region of x(0.2$<$x$<$0.6) which leads to an enhancement of nuclear structure function $F_{1,2}^A$~\cite{sajjadnpa}. 
 To obtain the contribution from the virtual mesons, we again evaluate lepton self energy and for this a diagram similar to the one shown in Fig.\ref{pse}
  is drawn, except that instead of a nucleon now there is a meson which 
 results in the change of a nucleon propagator by a meson propagator.
 This meson propagator does not correspond to the free mesons as one lepton (either electron or muon) can not decay into another lepton, one pion and X
 but corresponds to the mesons arising due to the nuclear medium 
 effects by using a modified meson propagator. These mesons are arising in the nuclear medium through 
 particle-hole(1p-1h), delta-hole(1$\Delta$-1h), 1p1h-1$\Delta$1h, 2p-2h,  etc. interactions as depicted in Fig.\ref{meson}. 
 
 In the present model, we have considered the contribution from 
 $\pi$ and $\rho$ meson clouds following the many body field theoretical approach as used in the case of bound nucleons. We shall make use of the imaginary part of the meson propagators instead of spectral function.
 In the case of pion following Ref.\cite{marco1996}, we replace  
 \[-2\pi\frac{M}{E(\mathbf{p})}S_{h}(p_0,\mathbf{p})~W^{\alpha \beta}_{N} (p, q)\] in 
 Eq.(\ref{had_ten151}) by \[ImD(p)\; \theta(p_0)\; 2 W^{\alpha \beta}_{\pi} (p, q)\]
 where $D(p)$ is the pion propagator in the nuclear medium given by 
 \begin{equation}\label{dpi}
D (p) = [ {p_0}^2 - {\bf {p}}\,^{2} - m^2_{\pi} - \Pi_{\pi} (p_0, {\bf p}) ]^{- 1}\,,
\end{equation}
with
\begin{equation}\label{pionSelfenergy}
\Pi_\pi=\frac{f^2/m_\pi^2 F^2(p){\bf {p}}\,^{2}\Pi^*}{1-f^2/m_\pi^2 V'_L\Pi^*}\,.
\end{equation}
Here, $F(p)=(\Lambda^2-m_\pi^2)/(\Lambda^2+{\bf {p}}\,^{2})$ is the $\pi NN$ form factor, $\Lambda$=1~$GeV$, $f=1.01$, $V'_L$ is
the longitudinal part of the spin-isospin interaction and $\Pi^*$ is the irreducible pion self energy that contains the 
contribution of particle - hole and delta - hole excitations.

Following a similar procedure, as done in the case of nucleon, the contribution of the pions to hadronic tensor in the nuclear medium may be written as \cite{marco1996}
\begin{equation}\label{W2pion}
W^{\mu \nu}_{A, \pi} = 3 \int d^3 r \; \int \frac{d^4 p}{(2 \pi)^4} \;
\theta (p_0) (- 2) \; Im D (p) \; 2 m_\pi W^{\mu \nu}_{\pi} (p, q)
\end{equation}
A factor of 3 arises due to the three charge state of pions and a factor of 2 is absent as compared to the nucleon as the pions are spinless particles.

Eq.(\ref{W2pion}) also contains the contribution of the pionic contents of the nucleon. Since these pionic contents are already contained in the
sea contribution of nucleon, therefore, the pionic 
contribution of the nucleon is to be subtracted from Eq.(\ref{W2pion}), in order to calculate the contribution from the 
pion excess in the nuclear medium. This is obtained by replacing $I m D (p)$ by $\delta I m D (p)$~\cite{marco1996} as
\begin{equation}
Im D (p) \; \rightarrow \; \delta I m D (p) \equiv I m D (p) - \rho \;
\frac{\partial Im D (p)}{\partial \rho} \left|_{\rho = 0} \right.
\end{equation}
which leads to
\begin{equation}\label{abcf1}  
F_{1, \pi}^A (x_\pi,Q^2) = - 6 A M \int  d^3 r   \int  \frac{d^4 p}{(2 \pi)^4} \theta (p_0) ~\delta I m D (p) \;2m_\pi~\left[\frac{F_{1\pi}(x_\pi,Q^2)}{m_\pi}
~+~\frac{{|{\bf p}|^2~-~p_{z}^2}}{2(p_0~q_0~-~p_z q_z)}\frac{F_{2\pi}(x_\pi,Q^2)}{m_\pi}\right] 
\end{equation}
where $x_\pi=-\frac{Q^2}{2p \cdot q}$.
\begin{figure}
\begin{center}
\includegraphics[height=0.3\textheight,width=0.75\textwidth]{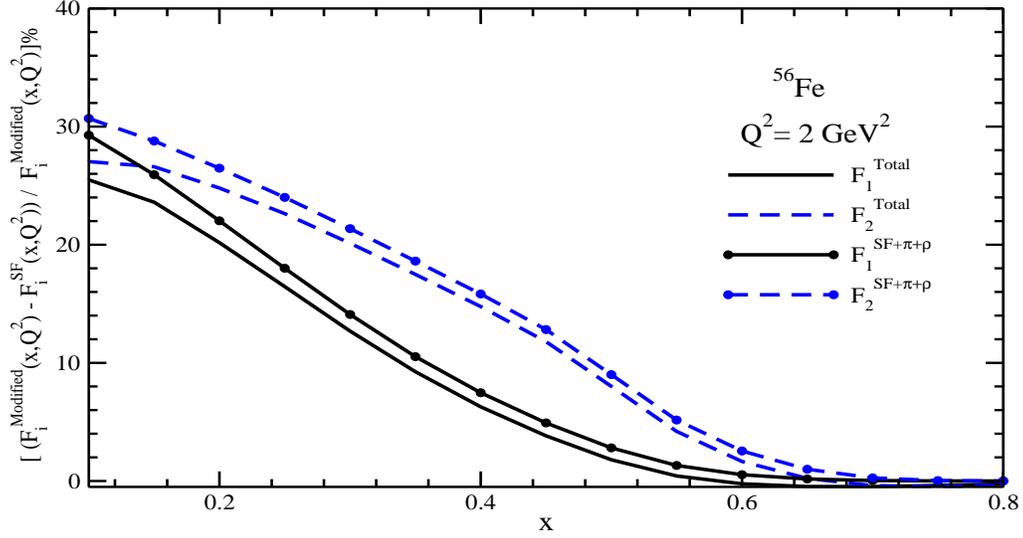}
\caption{$r_i={\frac{F_i^{Modified}(x,Q^2)~-~F_i^{SF}(x,Q^2)}{F_i^{Modified}(x,Q^2)}}$~(i=1,2) in $\%$ vs x, at $Q^2= 2 ~GeV^2$ in $^{56}Fe$. Here $F_i^{SF}$ stands for nuclear 
structure functions $F_{_{i}}^A(x,Q^2)~(i=1,2)$ obtained using spectral function only and $F_i^{Modified}$ stands for nuclear structure 
functions $F_{_{i}}^A(x,Q^2)~(i=1,2)$ evaluated (i) with mesonic effects along with the spectral function, and (ii) when shadowing is also included in (i).
The solid dotted(solid) line is 
the result obtained for $r_1$ using case i(ii) and dashed dotted(dashed) line is 
the result obtained for $r_2$ using case i(ii).}
    \label{Fignew1}
    \vspace{3mm}
\end{center}
\end{figure}

Following the same procedure as for $F_{2~N}^A(x,Q^2)$ obtained in Eq.(\ref{had_ten151}), 
the expression for $F_{2, \pi}^A (x)$, is given by
\begin{eqnarray} \label{pion_f21}
F_{_{2, \pi}}^A(x_\pi,Q^2)  &=&  -6 \int \, d^3 r \, \int \frac{d^4 p}{(2 \pi)^4} \, 
        \theta (p_0) ~\delta I m D (p) \;2m_\pi~\times \nonumber \\
&&\left[\frac{Q^2}{q_z^2}\left( \frac{|{\bf p}|^2~-~p_{z}^2}{2m_\pi^2}\right)  
+  \frac{(p_0~-~p_z~\gamma)^2}{m_\pi^2} \left(\frac{p_z~Q^2}{(p_0~-~p_z~\gamma) q_0 q_z}~+~1\right)^2\right]~\frac{m_\pi}{p_0~-~p_z~\gamma} ~F_{2\pi}(x_\pi,Q^2)
\end{eqnarray}
Similarly the contribution of the $\rho$-meson cloud to the structure function is taken into account in analogy with the above model and the
rho structure function is written as~\cite{marco1996}

\begin{eqnarray} \label{F2rho}
F_{1, \rho}^A(x_\rho,Q^2) = - 12 A M \int  d^3 r   \int  \frac{d^4 p}{(2 \pi)^4} \theta (p_0) ~\delta I m D_\rho (p) \;2m_\rho
~\left[\frac{F_{1\rho}(x_\rho,Q^2)}{m_\rho}~+~\frac{{|{\bf p}|^2~-~p_{z}^2}}{2(p_0~q_0~-~p_z~q_z)}
\frac{F_{2\rho}(x_\rho,Q^2)}{m_\rho}\right]
 \end{eqnarray}
 
\begin{eqnarray} \label{F2rho1}
F_{_{2, \rho}}^A(x_\rho,Q^2)  &=& -12 \int \, d^3 r \, \int \frac{d^4 p}{(2 \pi)^4} \, 
        \theta (p_0) ~\delta I m D_\rho (p) \;2m_\rho~\times \nonumber \\
&&\left[\frac{Q^2}{q_z^2}\left( \frac{|{\bf p}|^2~-~p_{z}^2}{2m_\rho^2}\right)  
+  \frac{(p_0~-~p_z~\gamma)^2}{m_\rho^2} \left(\frac{p_z~Q^2}{(p_0~-~p_z~\gamma) q_0 q_z}~+~1\right)^2\right]~\frac{m_\rho}{p_0~-~p_z~\gamma} ~F_{2\rho}(x_\rho,Q^2)~~~~
\end{eqnarray}

where $x_\rho=-\frac{Q^2}{2p \cdot q}$ and $D_{\rho} (p)$ is now the $\rho$-meson propagator in the medium given by:
\begin{equation}\label{dro}
D_{\rho} (p) = [ {p_0}^2 - {\bf{p}}\,^{2} - m^2_{\rho} - \Pi^*_{\rho} (p_0, {\bf p}) ]^{- 1}\,,
\end{equation}
where
\begin{equation}\label{rhoSelfenergy}
\Pi^*_\rho=\frac{f^2/m_\rho^2 C_\rho F_\rho^2(p){\bf{p}}\,^{2}\Pi^*}{1-f^2/m_\rho^2 V'_T\Pi^*}\,.
\end{equation}
Here, $V'_T$ is the transverse part of the spin-isospin interaction, $C_\rho=3.94$, $F_\rho(p)=(\Lambda_\rho^2-m_\rho^2)/(\Lambda_\rho^2+{\bf{p}}\,^{2})$ is the $\rho NN$ form factor, 
$\Lambda_\rho$=1~$GeV$, $f=1.01$, and $\Pi^*$ is the irreducible rho self energy that contains the contribution of particle - hole and delta - hole excitations. Quark and antiquark PDFs for pions have been taken from the parameterization 
given by Gluck et al.\cite{Gluck:1991ey} and for the 
rho mesons we have taken the same PDFs as for the pions. 

\begin{figure}
\begin{center} 
 \includegraphics[height=0.65\textheight,width=0.85\textwidth]{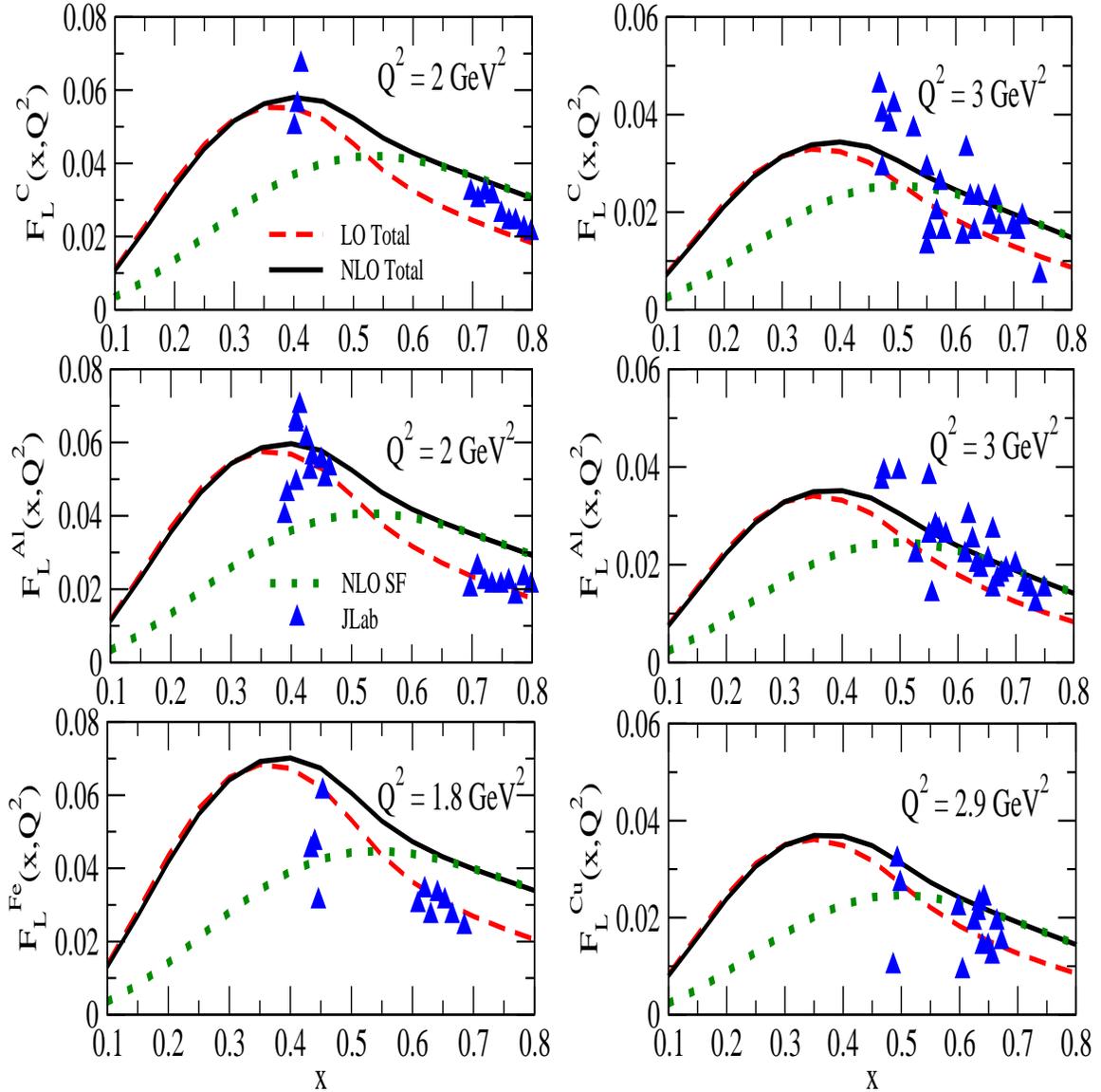}
 \caption{Results of the full calculation at LO(LO~Total) and NLO(NLO~Total) for the longitudinal structure function $F_L^A(x,Q^2)$ vs x, 
 in various nuclei($A$=$^{12}C,~^{27}Al,~^{56}Fe$~and~$^{63}Cu$). These results are presented at 
 different $Q^2$. Experimental points are JLab data~\cite{Mamyan:2012th}. 
  At NLO, these results are also presented with the spectral function(NLO SF). Nuclear targets $^{56}Fe$ and $^{63}Cu$ are treated as
 nonisoscalar in the present calculation.}
 \label{Figflq2var_all}
 \end{center}
 \end{figure}

 We have tested that our model fulfills the momentum
sum rule as expressed in Eq.(88) of Ref~\cite{Kulagin1989}  (including also the $\rho$ meson).
The procedure is straight forward by using Eqs. 36-37 from Kulagin and Petti and Eq.(10) from
our paper published in Ref.~\cite{sajjadnpa}. The pion $\bar y_\pi$ and nucleon $\bar y_N$ fractions of the light cone momentum are 
related by
\begin{equation}
 \bar y_\pi+ \bar y_N = \frac{M_A}{A M},
\end{equation}
where $M_A$ is the nucleus mass. The nucleon quantities can be easily obtained from the spectral function.
 Our results are like this: For iron $\bar y _N= 0.967$, $\pi$+$\rho$ should account for 0.024. 

\section{Results and Discussion} \label{sec:RE}
For the numerical calculations we have used Eq.(\ref{conv_WA1}) for $F_{1~N}^A(x,Q^2)$ and Eq.(\ref{had_ten151}) for $F_{2~N}^A(x,Q^2)$. The 
contributions of $\pi$ and $\rho$ mesons have been taken into account using Eq.(\ref{abcf1}) for $F_{1~\pi}^A(x,Q^2)$, 
Eq.(\ref{pion_f21}) for $F_{2~\pi}^A(x,Q^2)$, Eq.(\ref{F2rho}) for $F_{1~\rho}^A(x,Q^2)$ and Eq.(\ref{F2rho1})
for $F_{2~\rho}^A(x,Q^2)$. We have added the mesonic contributions to the nucleonic contribution to get the full result.
It should be mentioned that all the parameters entering in the above mentioned equations have been fixed by
 our earlier~\cite{sajjadnpa,prc84} works and have not been taken as free parameters. We have used CTEQ parameterization~\cite{cteq} 
 for the quark/antiquark PDFs for nucleon and the quark/antiquark
parameterization for pion from the works of Gluck et al.~\cite{Gluck:1991ey}. For the rho meson we have taken the same
PDFs as for the pion. Moreover, while determining nucleon structure functions in terms of parton 
distribution functions, it is important to include target mass correction(TMC) to take into account the effects associated with the non-zero mass of the target. 
 TMC effects are more pronounced 
at large x and moderate $Q^2$. We have taken into account the target mass correction following the works of Schienbein et al.~\cite{Schienbein:2007gr}.  The 
shadowing effect has been incorporated following Ref.~\cite{Petti} and is pronounced at low x(x$<$0.1). Therefore, in the present studied 
 region of x($>0.1$) the effect of shadowing is almost negligible and the results for this are not explicitly shown in all the figures. 

 The results are presented for different cases. The first case is when we take the contribution 
 of nucleon spectral function only i.e.
 Eq.(\ref{conv_WA1}) for $F_{1~N}^A(x,Q^2)$ and Eq.(\ref{had_ten151}) for $F_{2~N}^A(x,Q^2)$, 
 which includes the effect of Fermi motion, 
nuclear binding and nucleon correlations, and perform the calculation at the
leading order(LO). The numerical results are referred as LO(SF). We then include the contribution from
meson clouds as well as shadowing effect and this we call as the results of our full model. This we refer as LO(Total). These results have also 
been obtained at NLO for both the cases. The results in these cases are referred as NLO(SF) and NLO(Total), respectively.

In Fig.\ref{FigF2_all}, we have presented the results for $F_{2}^A(x,Q^2)$ in various nuclear targets like $^{12}C$, $^{27}Al$, $^{56}Fe$ and $^{63}Cu$
 at different $Q^2$. We find that in $^{12}C$ at $Q^2=2~GeV^2$, the results obtained using full model at LO, 
is about 18$\%$ higher at x=0.2 in comparison to
 the results obtained using spectral function only. This difference decreases with the increase in x, for example at x=0.4, it is 12$\%$ and becomes
 almost negligible at x=0.6. We observe that this difference increases with the increase in mass number i.e. for  $^{27}Al$, it is $\sim$ 20$\%$ higher 
 at x=0.2 and 14$\%$ at x=0.4, while in $^{56}Fe$ it is $\sim$ 23$\%$ higher 
 at x=0.2 and 17$\%$ at x=0.4. In $^{118}Sn$(not shown in the figure) this difference becomes about 25$\%$ at x=0.2 and 18$\%$ at x=0.4, whereas in $^{208}Pb$
 (not shown in the figure) this difference increases to 26$\%$ at x=0.2 and 20$\%$ at x=0.4. The difference is mainly due to mesonic effects and is negligible
  for $x\geq0.6$ for the nuclei considered here.  When the results obtained by using the full 
  model at NLO are compared with the results evaluated at LO, we find that the results decrease
  from the LO values. For example, in the case of $^{12}C$, it is lower by
   10$\%$ at x=0.2 and 5$\%$ at x=0.4. Then there is a cross over around x=0.5 and beyond that it increases with increase in x. For example, 
   it is around 35-38$\%$ larger at x=0.8 for the nuclei considered here. The effect of shadowing has been found
   to be almost negligible in the presently studied region of x. For example, 
   the effect of shadowing is around 1-2$\%$ at x$\sim$0.2 in $^{12}C$ which increases to 2-3$\%$ for heavier nuclei like $^{118}Sn$ and $^{208}Pb$. 
     We have also shown in this figure, JLab experimental data~\cite{Mamyan:2012th} and find that our results obtained with full model at 
  NLO agree reasonably well with the JLab data.

In Fig.\ref{Figq2var_all}, we present the results for $2xF_{1}^A(x,Q^2)$ obtained using Eqs.(\ref{conv_WA1}), (\ref{abcf1}) and (\ref{F2rho}). 
The results are presented with full model at LO and NLO in
several nuclei like $^{12}C$, $^{27}Al$, $^{56}Fe$ and $^{63}Cu$  
at various values of $Q^2$. We find that in general these results are qualitatively similar in nature to that found in the case of $F_2^A(x,Q^2)$, 
however, quantitatively
there is some variation, specially in the region of low x where mesonic effects play a role. The results are also 
 compared with the JLab data~\cite{Mamyan:2012th} and we find a reasonable agreement with the experimental data. 

We separately show the effect of mesonic contribution and shadowing effect in $F_{1}^A(x,Q^2)$ and $F_{2}^A(x,Q^2)$, by 
presenting the results for $r_i={\frac{F_i^{Modified}(x,Q^2)~-~F_i^{SF}(x,Q^2)}{F_i^{Modified}(x,Q^2)}}$,(i=1,2) in Fig.\ref{Fignew1}.
These results are shown in $^{56}Fe$, where
 $F_i^{SF}(x,Q^2)$ stands for the results obtained for the nuclear structure functions using the spectral function 
 only while $F_i^{Modified}(x,Q^2)$ is the result obtained 
when  we include (i) mesonic($\pi+\rho$) 
contributions and (ii)  mesonic($\pi+\rho$) contributions and shadowing effects. At low $Q^2$, we find that the mesonic contributions are larger which become 
 smaller with the increase in $Q^2$. For example, at $Q^2=2~GeV^2$, the mesonic contribution is 24-28$\%$ at x=0.2, 
which becomes 2-4$\%$ at x=0.6. Similarly at $Q^2=5~GeV^2$(not shown in the figure), the mesonic contribution is found to be 20-24$\%$ at x=0.2, 
which becomes 1-2$\%$ at x=0.6. We also find that the mesonic contributions to $F_{2}^A(x,Q^2)$ at a given x and $Q^2$ is larger than the 
mesonic contribution in $F_{1}^A(x,Q^2)$ over the whole range of x and $Q^2$. The results presented in Figs.\ref{FigF2_all} and \ref{Figq2var_all} 
 for $F_{2}^A(x,Q^2)$ and $2xF_{1}^A(x,Q^2)$ give useful predictions for $F_{1,2}^A(x,Q^2)$ at low x, which may be tested 
 in future experiments. It may be observed that the mesonic contributions are really different for $F_{1}^A(x,Q^2)$ and $F_{2}^A(x,Q^2)$ 
 structure functions and when shadowing contributions are included both curves get reduced because the 
 shadowing goes in the opposite direction to the enhancement due to meson cloud contribution and this can be viewed as a kind of compensating effect.
  
   In Fig.\ref{Figflq2var_all}, we present the results for the longitudinal structure function $F_L^A(x,Q^2)$ using the 
   full model at LO as well as at NLO. Here we have also shown the results  
with the spectral function only, obtained at NLO. These results are presented for various nuclear targets 
like $^{12}C$, $^{27}Al$, $^{56}Fe$ and $^{63}Cu$ at different $Q^2$.  
We find that mesonic contributions to $F_L^A(x,Q^2)$ are quite large at low x and $Q^2$ and becomes negligible for $x>0.6$
for almost all values of $Q^2$ studied here. For example, in $^{12}C$ at $Q^2= 2~GeV^2$, the mesonic contributions is $\sim 58\%$ (of total $F_L^A(x,Q^2)$)
at x=0.2 which reduces to $20\%$ at x=0.5. The mesonic contributions 
increases for all x(x $<$ 0.6) with the increase in mass number as we go from $^{12}C$ to $^{63}Cu$ as shown in Fig.\ref{Figflq2var_all}.
 Thus the results presented in this figure are also an equivalent proof of the fact that the mesonic contributions to $F_{1}^A(x,Q^2)$ and $F_{2}^A(x,Q^2)$
 are really different. This is because, in principle, $F_{L}^A(x,Q^2)$ should be zero if 
 Callan-Gross relationship were exactly fulfilled. 
 While the nucleonic contributions(SF only) of $F_2$ and $2xF_1$ show a more prominent trend to cancel while computing $F_L$, whereas the mesonic contributions
  when added to it leads to a significant behaviorial change which results in a broken Callan-Gross relation(compare the curve shown by the dotted lines to the curve
  shown by the solid line in Fig.\ref{Figflq2var_all}). 
The results at NLO are larger than LO for $x>0.4$ and are in better agreement with the experimental results for $F_L^A(x,Q^2)$ for JLab~\cite{Mamyan:2012th}.
However, there are some exceptions specially at low $Q^2$ ($\leq 2~GeV^2$) as seen in the case of $^{27}$Al and $^{56}$Fe. At smaller $Q^2$ in the region of
$Q^2\leq 2~GeV^2$, the non-perturbative QCD effects and their possible enhancement in nuclear medium may play an important role which 
are beyond the scope of this work. However the present work makes important prediction for $F_L^A(x,Q^2)$ in lower x region which may be tested in
future experiments.

\begin{figure}
\begin{center}
\includegraphics[height=0.35\textheight,width=0.75\textwidth]{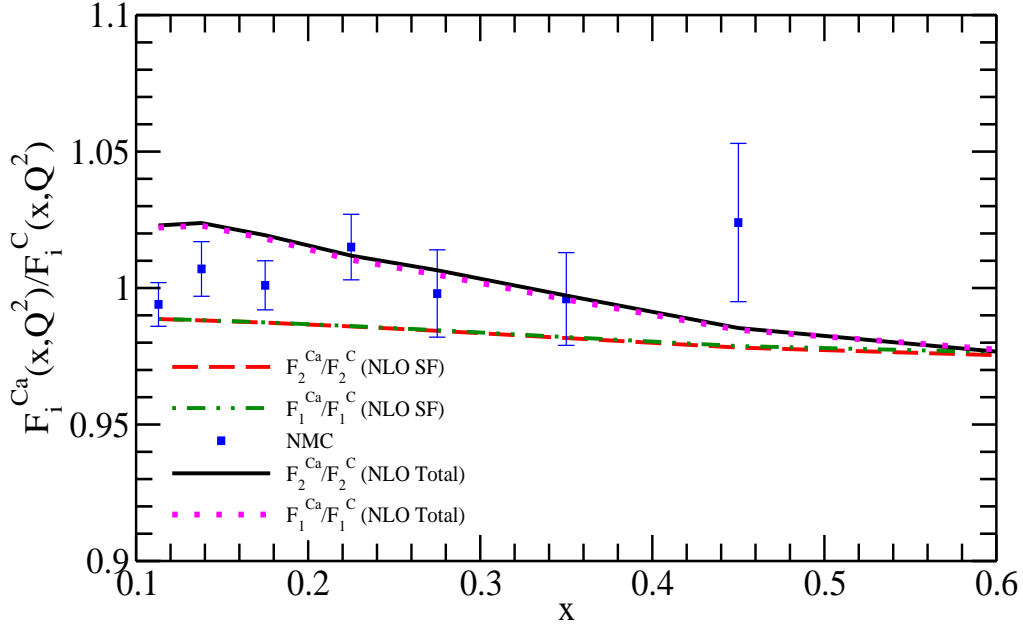}
\caption{Results obtained using the spectral function(NLO~SF) and the full model(NLO~Total) at NLO for the ratio 
$\frac{F_i^{Ca}(x,Q^2)}{F_i^{C}(x,Q^2)}$;~i=1,2. Experimental points are 
the NMC result~\cite{nmcca2d}.}
    \label{Fig118Sn}
    \vspace{3mm}
\end{center}
\end{figure}

We present the results for $\frac{F_1^{Ca}(x,Q^2)}{F_1^C(x,Q^2)}$ and 
$\frac{F_2^{Ca}(x,Q^2)}{F_2^C(x,Q^2)}$ in Fig.\ref{Fig118Sn}, using the spectral function and the full model 
at NLO and compared the results with NMC data~\cite{nmcca2d}. We find that the results are in better agreement with the experimental 
observations when mesonic contributions are included in addition to the nuclear medium effects. We find a strong 
nuclear dependence as the absolute values of the  slope increases with the 
increase in the mass number. This may be noted that 
 although mesonic cloud contributions seem to be very different for $F_{1}^A(x,Q^2)$ and $F_{2}^A(x,Q^2)$ separately, 
 they also seem to cancel when computing the ratios between structure functions for different nuclei.
 \section{Conclusion} \label{sec:Summary}
To conclude, we have studied nuclear medium effects in electromagnetic nuclear structure 
functions $F_{1}^A(x,Q^2)$ and $F_{2}^A(x,Q^2)$, and the 
 longitudinal structure function $F_L^A(x,Q^2)$. For the nuclear medium effects, 
we took into account 
Fermi motion, nuclear binding, nucleon correlations, effect of meson degrees of freedom, etc. The calculations are performed 
both at LO and NLO.
  
 The theoretical expressions for $F_1^A(x,Q^2)$ and $F_2^A(x,Q^2)$ have been obtained without assuming Callan-Gross
 relation at nuclear level but it has been assumed at nucleonic(and mesonic) level while computing $F_{1,2}^N(x,Q^2)$ and $F_{1,2}^{\pi,~\rho}(x,Q^2)$
 for free nucleons and mesons. The theoretical results are then compared with the experimental data for 
 $F_{1}^A(x,Q^2)$, $F_{2}^A(x,Q^2)$, $F_{L}^A(x,Q^2)$ from  JLab data~\cite{Mamyan:2012th} and found in good agreement with them (for x $>$ 0.4) 
 except for $F_{L}^A$ at very low $Q^2 < 2 ~GeV^2$. The present work makes predictions for nucleon electromagnetic structure
 functions at lower x in various nuclei in the region of $2~<~Q^2~<~4GeV^2$,  which will be useful in analyzing the future experiments being
 done for studying the nuclear medium effects in lepton nucleus scattering at low and moderate $Q^2$.
 The results are also compared with the old NMC data on $\frac{F_{1,2}^{Ca}(x,Q^2)}{F_{1,2}^{C}(x,Q^2)}$ with good agreement and predictions
 are made for other nuclei.
 \begin{acknowledgments}
  M. S. A. is thankful to Department of Science
and Technology(DST), Government of India for providing financial assistance under Grant No. SR/S2/HEP-18/2012. 
I.R.S. thanks University of Granada for financial support through 'Programa de Fortalecimiento de I+D.  
 \end{acknowledgments}

 \appendix
 \section{Properties of the spectral function}\label{app}
 The hole and particle spectral functions fulfill the following relations,
\begin{equation}
 \int_{-\infty}^\mu\,dp_0\;S_h(p_0,\mathbf{p})=n(\mathbf{p})
\end{equation}

\begin{equation}
 \int^{\infty}_\mu\,dp_0\;S_p(p_0,\mathbf{p})=1-n(\mathbf{p})
\end{equation}
where $n(\mathbf{p})$ is the Fermi occupation number.

Therefore, the following sum rule is also full filled
\begin{equation}
 \int_{-\infty}^\mu\,dp_0\;S_h(p_0,\mathbf{p})+
 \int^{\infty}_\mu\,dp_0\;S_p(p_0,\mathbf{p})=1
\end{equation}

In the absence of 
interactions, the nucleon energy $p_0$ is the free relativistic energy $E(\mathbf{p})$ and the dressed propagator $G(p)$ 
reduces to the free propagator $G^0(p)$ i.e. if $\Sigma^N(p)=0$ then 
\begin{equation}
 S_h(p_0,\mathbf{p})=S_p(p_0,\mathbf{p})=\delta(p_0-E(\mathbf{p}))
\end{equation}
then
\begin{eqnarray}
 \int_{-\infty}^\mu\,dp_0\;S_h(p_0,\mathbf{p})&=&
 \int_{-\infty}^\mu\,dp_0\;\delta(p_0-E(\mathbf{p}))=
 \left\lbrace
\begin{tabular}{ll}
1 & $\textrm{if}\quad \mu>E(\mathbf{p})$\\
0 & $\textrm{if}\quad \mu<E(\mathbf{p})$
\end{tabular}
\right.\\
 \int^{\infty}_\mu\,dp_0\;S_p(p_0,\mathbf{p})&=&
 \int^{\infty}_\mu\,dp_0\;\delta(p_0-E(\mathbf{p}))=
 \left\lbrace
\begin{tabular}{ll}
1 & $\textrm{if}\quad \mu<E(\mathbf{p})$\\
0 & $\textrm{if}\quad \mu>E(\mathbf{p})$
\end{tabular}
\right.
\end{eqnarray}

Thus in the limiting case of vanishing self energy the expression for the spectral functions given in Eqs.\ref{sh} and \ref{sp} collapse to a representation of 
Dirac delta function. 
If $E(\mathbf{p})$ is the total relativistic energy, then $\mu$ must have the 
nucleon mass $M$ incorporated i.e.
\begin{equation}
 \mu=M+\epsilon_F
\end{equation}
With this definition, we can perform a constant shift in the integration variable $p_0$, 
given by:
\begin{equation}
 p_0=\omega+M
\end{equation}
And with this shift, the integrals stand:
\begin{eqnarray*}
 \int_{-\infty}^\mu\,dp_0\;S_h(p_0,\mathbf{p})&=&
 \int_{-\infty}^{\mu-M}\,d\omega\;\delta(\omega+M-E(\mathbf{p}))=
 \left\lbrace
\begin{tabular}{ll}
1 & $\textrm{if}\quad \mu-M>E(\mathbf{p})-M\Rightarrow \epsilon_F>\epsilon(\mathbf{p})$\\
0 & $\textrm{if}\quad \mu-M<E(\mathbf{p})-M\Rightarrow \epsilon_F<\epsilon(\mathbf{p})$
\end{tabular}
\right.\\
 \int^{\infty}_\mu\,dp_0\;S_p(p_0,\mathbf{p})&=&
 \int^{\infty}_{\mu-M}\,d\omega\;\delta(\omega+M-E(\mathbf{p}))=
 \left\lbrace
\begin{tabular}{ll}
1 & $\textrm{if}\quad \mu-M<E(\mathbf{p})-M\Rightarrow\epsilon_F<\epsilon(\mathbf{p})$\\
0 & $\textrm{if}\quad \mu-M>E(\mathbf{p})-M\Rightarrow\epsilon_F>\epsilon(\mathbf{p})$
\end{tabular}
\right.
\end{eqnarray*}
where $\epsilon(\mathbf{p})=E(\mathbf{p})-M$ is the nucleon kinetic energy, which
in the non-relativistic regime can be approximated by
\begin{equation}
 \epsilon(\mathbf{p})\approxeq\frac{\mathbf{p}^2}{2M}
\end{equation}
If $\epsilon_F=\frac{p^2_F}{2M}$ is the Fermi energy, then the two step functions
$\theta(p_F-|\mathbf{p}|)$ and $\theta(|\mathbf{p}|-p_F)$ are the solutions of the above
integrals, namely
\begin{eqnarray}
 \int_{-\infty}^\mu\,dp_0\;S_h(p_0,\mathbf{p})&=&
 \theta(p_F-|\mathbf{p}|)\equiv n_0(\mathbf{p})\\
 \int^{\infty}_\mu\,dp_0\;S_p(p_0,\mathbf{p})&=&
\theta(|\mathbf{p}|-p_F)\equiv 1-n_0(\mathbf{p})
\end{eqnarray}
Thus in the absence of interactions, the full dressed propagator reduces to free one and one may write
\begin{eqnarray}
 G(p_0,\mathbf{p})&=&\frac{M}{E(\mathbf{p})}\sum_{r}u_r(\mathbf{p})\bar{u}_r(\mathbf{p})
 \left[\int_{-\infty}^\mu d\omega\;\frac{S_h(\omega,\mathbf{p})}{p_0-\omega-i\eta}
 +\int_{\mu}^{\infty} d\omega\;\frac{S_p(\omega,\mathbf{p})}{p_0-\omega+i\eta}\right]\nonumber\\
 &=&\frac{M}{E(\mathbf{p})}\sum_{r}u_r(\mathbf{p})\bar{u}_r(\mathbf{p})
  \left[\int_{-\infty}^\mu d\omega\;
  \frac{\delta\left(\omega-E(\mathbf{p})\right)}{p_0-\omega-i\eta}
 +\int_{\mu}^{\infty} d\omega\;
 \frac{\delta\left(\omega-E(\mathbf{p})\right)}{p_0-\omega+i\eta}\right]\nonumber\\
 &=&\frac{M}{E(\mathbf{p})}\sum_{r}u_r(\mathbf{p})\bar{u}_r(\mathbf{p})
  \left[\frac{\theta(p_F-|\mathbf{p}|)}{p_0-E(\mathbf{p})-i\eta}
 +\frac{\theta(|\mathbf{p}|-p_F)}{p_0-E(\mathbf{p})+i\eta}\right]\nonumber\\
 &=&\frac{M}{E(\mathbf{p})}\sum_{r}u_r(\mathbf{p})\bar{u}_r(\mathbf{p})
  \left[\frac{n_0(\mathbf{p})}{p_0-E(\mathbf{p})-i\eta}
 +\frac{1-n_0(\mathbf{p})}{p_0-E(\mathbf{p})+i\eta}\right]
\end{eqnarray}

The hole spectral function $S_h (p_0, {\bf p})$ is physically interpreted as equal to the joint
probability of (i) removing of a nucleon with momentum ${\bf p}$ 
 from the correlated 
  ground state, and (ii) of finding the resulting system of (A-1) nucleons with an energy in the interval $p_0$ and $p_0 + dp_0$.

 The normalization of this spectral function is obtained by imposing the baryon number conservation following
 Ref.~\cite{Frankfurt:1985ui}:
\begin{eqnarray}\label{NNmu}
\left<N|B^{\mu}|N\right>\equiv \bar u({{\bf p}}) \gamma^{\mu} u({{\bf p}})~=~B \frac{p^{\mu}}{M};~ B=1,~ p^{\mu}\equiv(E({{\bf p}}),{{\bf p}})
\end{eqnarray}
and
\begin{eqnarray}\label{AAmu}
\left<A|B^{\mu}|A\right>=-\int \frac{d^4p}{(2\pi)^4}V i Tr[G(p_0,{{\bf p}})\gamma^\mu]e^{ip_0\eta}.
\end{eqnarray}
where $V$ is the volume of the normalization box and $exp({ip_0\eta})$, with $\eta \rightarrow 0^+$, is the convergence factor for loops 
appearing at equal times.

Using the expression given by Eq.(\ref{Gp}), it can be seen that the convergence factor limits the contribution to the hole spectral 
function and one gets
\begin{eqnarray} \label{ABmu}
\left<A|B^{\mu}|A\right>&=&V\int\frac{d^{3}p}{(2\pi)^3}\frac{M}{E({{\bf p}})}{ Tr} \left[\sum_{r}u_{r}({{\bf p}})\bar u_{r}({{\bf p}})\gamma^{\mu}\right]\int_{-\infty}^{\mu}S_{h}(\omega,{\bf{p}})d\omega \nonumber\\
&&=V\int\frac{d^{3}p}{(2\pi)^3}\frac{M}{E({{\bf p}})}{ Tr} \left[\frac{(\not p+M)_{on shell}}{2M}\gamma^{\mu}\right]\int_{-\infty}^{\mu}S_{h}(\omega,{\bf{p}})d\omega \nonumber\\
&&=2V\int\frac{d^{3}p}{(2\pi)^3}\frac{M}{E({{\bf p}})}\frac{p^{\mu}_{on shell}}{M}\int_{-\infty}^{\mu}S_{h}(\omega,{\bf{p}})d\omega\equiv B\frac{p_A^\mu}{M_A} 
\end{eqnarray}
It is to be noted that in the last step we have imposed that this matrix element gives the right current with $B$ baryons, in 
analogy to the expression given by Eq.(\ref{NNmu}). $p_A^\mu$ is the momentum of the nucleus. The operator $({\not p} + M)$ comes 
from $\sum u_r({{\bf p}})\bar u_r({{\bf p}})$ which depends only on ${{\bf p}}$, 
and that corresponds to a free particle with $p^\mu=(E({{\bf p}}),{{\bf p}})$, therefore, the operator $(\not p + M)$ when the
particle is on shell is written as 
\begin{eqnarray} \label{norm1}
2V\int\frac{d^{3}p}{(2\pi)^{3}}\int_{-\infty}^{\mu}S_{h}(\omega,{\bf{p}}) d\omega= B~=1
\end{eqnarray}
We have ensured that the spectral function is properly normalized and checked it by
obtaining the correct baryon number and binding energy for a given nucleus.
In the local density approximation, the spectral
functions of protons and neutrons are the function of local Fermi momentum.
The equivalent normalization to Eq.(\ref{norm1}) is written as
\begin{eqnarray} \label{norm2}
2\int\frac{d^{3}p}{(2\pi)^{3}}\int_{-\infty}^{\mu}S_{h}(\omega,p,p_{F_{p,n}}({\bf {r}})) d\omega= \rho_{p,n}({\bf {r}})
\end{eqnarray}
where factor 2 is due to two possible projections of spin $\frac{1}{2}$ particle. $p_{F_{p(n)}}$ is the Fermi momentum of
proton(neutron) inside the nucleus which is 
expressed in terms of proton(neutron) densities given by $p_{F_{p(n)}}({\bf
r})= \left[ 3\pi^{2} \rho_{p(n)}({\bf {r}}) \right]^{1/3}$. These nucleon densities are in turn related with the nuclear densities 
$\rho(r)$(like $\rho_p(r)=\frac{Z}{A}\rho(r)$ and $\rho_n(r)=\frac{A-Z}{A}\rho(r))$, the parameters of which are determined 
from lepton scattering experiments. In the present calculation we have used harmonic
 oscillator density for $^{12}C$ nucleus and two  parameter Fermi density for $^{27}Al$,  $^{56}Fe$, $^{63}Cu$, $^{118}Sn$, 
 $^{197}Au$, and $^{208}Pb$ nuclei which are taken
 from Refs.~\cite{De Jager:1987qc}-\cite{GarciaRecio}.
This leads to the normalization condition individually satisfied by proton and neutron as
\begin{eqnarray}\label{norm4}
2 \int d^3 r \;  \int \frac{d^3 p}{(2 \pi)^3} \int^{\mu}_{- \infty} \; S_h (\omega, {\bf{p}}, \rho_p(r)) \; d \omega &=& Z\,,\nonumber\\
2 \int d^3 r \;  \int \frac{d^3 p}{(2 \pi)^3} \int^{\mu}_{- \infty} \; S_h (\omega, {\bf {p}}, \rho_n(r)) \; d \omega &=& A-Z\,,
\end{eqnarray}
For a symmetric nuclear matter of density $\rho({\bf r})$, there is a unique Fermi momentum given by 
$p_{F}({\bf {r}})= \left[ 3\pi^{2} \rho({\bf {r}})/2\right]^{1/3}$ for which we obtain
\begin{eqnarray} \label{norm3}
4\int\frac{d^{3}p}{(2\pi)^{3}}\int_{-\infty}^{\mu}S_{h}(\omega,p,p_{F}({\bf {r}})) d\omega= \rho({\bf {r}})
\end{eqnarray}
leading to the normalization condition given by
\begin{equation}\label{norm5}
4 \int d^3 r \;  \int \frac{d^3 p}{(2 \pi)^3} 
\int^{\mu}_{- \infty} \; S_h (\omega, {\bf{p}}, \rho(r)) 
\; d \omega = A\,,
\end{equation}
where $\rho(r)$ is the baryon density for the nucleus which is normalized to $A$ and is taken from the lepton nucleus scattering 
experiments.
Also we calculate the average kinetic and total nucleon energy given by~\cite{marco1996}: 
\begin{eqnarray}
<T>= \frac{4}{A} \int d^3 r \;  \int \frac{d^3 p}{(2 \pi)^3} (E({\bf p})-M) 
\int^{\mu}_{- \infty} \; S_h (p_0, {\bf p}, \rho(r)) 
\; d p_0\,,
\end{eqnarray}
\begin{eqnarray}
<E>= \frac{4}{A} \int d^3 r \;  \int \frac{d^3 p}{(2 \pi)^3}  
\int^{\mu}_{- \infty} \; S_h (p_0, {\bf p}, \rho(r)) 
\; p_0 d p_0\,,
\end{eqnarray}
and the binding energy per nucleon given by~\cite{marco1996}:
\begin{equation}
|E_A|=-\frac{1}{2}(<E-M>+\frac{A-2}{A-1}<T>)
\end{equation}
Here we have ensured that we retrieve the kinetic energy $<T>$ and the total energy $<E>$ for the nucleon. 
We have tabulated in Table~\ref{table1}, the kinetic energy per nucleon and the binding energies for nuclei used in the numerical calculations.
\begin{table}
\begin{center}
\begin{tabular}{|c|c|c|c|}\hline\hline
Nucleus  & $< T >/A$ (MeV) & $B.E./A$ (MeV)\\
 \hline
$^{12}C$ & 20.2 & 7.6  \\\hline
$^{27}Al$&27.8& 8.3  \\\hline
$^{56}Fe$ & 30.0 & 8.8  \\\hline
$^{63}Cu$&29.3& 8.7  \\\hline
$^{118}Sn$&31.8&8.6  \\\hline
$^{197}Au$&33.7& 7.9 \\\hline
$^{208}Pb$&32.7&7.8  \\\hline\hline
\end{tabular}
\end{center}
\caption{ Kinetic energy per nucleon($< T >/A$) and binding energy per nucleon($B.E./A$) for $^{12}C$,
$^{27}Al$, $^{56}Fe$, $^{63}Cu$, $^{118}Sn$, $^{197}Au$,  and $^{208}Pb$.}
\label{table1}
\end{table}

\end{document}